\documentclass[11pt,a4paper]{article}
\pdfoutput=1

\usepackage{my_jcappub}
\usepackage{mathrsfs}
\usepackage{bbm}
\usepackage{bm}
\usepackage{dsfont}
\usepackage{tensor}
\usepackage{xspace}
\usepackage{aeguill}
\usepackage{wasysym}
\usepackage{microtype}
\usepackage{empheq}
\usepackage{ifthen}
\usepackage{subcaption}

\newcommand{\dd}{{\rm d}}
\newcommand{\pd}[3][]{\frac{\partial^{#1} #2}{\partial {#3}^{#1}}}
\newcommand{\ddf}[3][]{\frac{\dd^{#1} #2}{\dd {#3}^{#1}}}
\newcommand{\ph}{\varphi}
\newcommand{\eps}{\varepsilon}
\newcommand{\thet}{\vartheta}
\renewcommand{\i}{\mathrm{i}}
\newcommand{\define}{\equiv}
\newcommand{\delimiters}[4][]{
\ifthenelse{ \equal{#1}{1} }{  #2 #3 #4  }
					{ \ifthenelse{\equal{#1}{2}}{ \big#2 #3 \big#4 }
						{ \ifthenelse{\equal{#1}{3}}{ \Big#2 #3 \Big#4 }
							{ \ifthenelse{\equal{#1}{4}}{ \bigg#2 #3 \bigg#4 }
								{ \ifthenelse{\equal{#1}{5}}{ \Bigg#2 #3 \Bigg#4 }
									{ \left#2 #3 \right#4 }
								}
							}
						}
					}
													}

\newcommand{\pa}[2][]{\delimiters[#1]{(}{#2}{)}}
\newcommand{\pac}[2][]{\delimiters[#1]{[}{#2}{]}}
\newcommand{\paac}[2][]{\delimiters[#1]{\{}{#2}{\}}}
\newcommand{\abs}[2][]{\delimiters[#1]{|}{#2}{|}}
\newcommand{\ev}[2][]{\delimiters[#1]{\langle}{#2}{\rangle}}

\newcommand{\e}[1]{_\text{#1}}
\newcommand{\h}[1]{^\text{#1}}
\newcommand{\ex}[1]{\mathrm{e}^{#1}}
\newcommand{\U}[1]{\,\mathrm{#1}}

\newcommand{\vect}[1]{\boldsymbol{#1}}
\newcommand{\nul}[1]{#1}
\newcommand{\tr}{\mathrm{tr}}
\newcommand{\transpose}[1]{{#1}^{\rm T}}

\newcommand{\obs}{\text{o}}
\newcommand{\source}{\text{s}}
\newcommand{\prob}{\mathrm{Prob}}

\newcommand{\identity}{\boldsymbol{1}}
\newcommand{\zero}{\boldsymbol{0}}
\newcommand{\separation}{\xi}
\newcommand{\optical}{\mathcal{R}}
\newcommand{\jacobi}{\mathcal{D}}
\newcommand{\amplification}{\mathcal{A}}

\newcommand{\deformation}{\mathcal{S}}
\newcommand{\Weylmat}{\mathcal{W}}
\newcommand{\diffusion}{\mathcal{Q}}

\newcommand{\Ricfoc}{\mathscr{R}}
\newcommand{\Weylfoc}{\mathscr{W}}
\newcommand{\varDA}{\mathrm{var}(D\e{A})}

\newenvironment{system}{\left\{\begin{aligned}}{\end{aligned}\right.}

\title{The theory of stochastic cosmological lensing}

\author[a,b]{Pierre Fleury,}
\author[c]{Julien Larena,}
\author[a,b]{Jean-Philippe Uzan}

\affiliation[a]{Institut d'Astrophysique de Paris, UMR 7095 du CNRS, 98 bis Bd Arago, 75014 Paris, France.}
\affiliation[b]{Sorbonne Universit\'es, Institut Lagrange de Paris, 98 bis, Bd Arago, 75014 Paris, France.}
\affiliation[c]{Department of Mathematics, Rhodes University, Grahamstown 6140, South Africa}

\emailAdd{fleury@iap.fr}
\emailAdd{j.larena@ru.ac.za}
\emailAdd{uzan@iap.fr}

\abstract{On the scale of the light beams subtended by small sources, e.g. supernovae, matter cannot be accurately described as a fluid, which questions the applicability of standard cosmic lensing to those cases. In this article, we propose a new formalism to deal with small-scale lensing as a diffusion process: the Sachs and Jacobi equations governing the propagation of narrow light beams are treated as Langevin equations. We derive the associated Fokker-Planck-Kolmogorov equations, and use them to deduce general analytical results on the mean and dispersion of the angular distance. This formalism is applied to random Einstein-Straus Swiss-cheese models, allowing us to: (1) show an explicit example of the involved calculations; (2) check the validity of the method against both ray-tracing simulations and direct numerical integration of the Langevin equation. As a byproduct, we obtain a post-Kantowski-Dyer-Roeder approximation, accounting for the effect of tidal distortions on the angular distance, in excellent agreement with numerical results. Besides, the dispersion of the angular distance is correctly reproduced in some regimes.}

\arxivnumber{1508.07903}
\newcommand{\publishedID}{JCAP 11 (2015) 022}

\date{November 11, 2015}

\begin{document}
\maketitle
\flushbottom

\section{Introduction}\label{section1}

The understanding of light propagation in the Universe, in particular through the relation between distances and redshifts, is central for the interpretation of almost all cosmological observations. The standard approach consists in assuming that light propagates through a strictly homogeneous and isotropic Friedmann-Lema\^{\i}tre (FL) spacetime~\cite{pubook}, assumed to be a good model on cosmological scales.\footnote{See however Refs.~\cite{2014CQGra..31w4003G,2015arXiv150507800B,2015arXiv150606452G} for a recent debate on this specific issue.}
Such a crude---but surprisingly efficient---approximation can be refined by taking into account: (i)~the actual non-comobility of both the light sources and the observer; (ii)~the gravitational lensing caused by the large-scale structure. This more realistic description generally relies on the cosmological perturbation theory~\cite{Futamase:1989hba,Cooray:2005yp,Dodelson:2005zt}. At first order, it essentially introduces a dispersion of the distance-redshift relation with respect to the background FL prediction~\cite{Valageas:1999ch,Bonvin:2005ps,Meures:2011gp,2013JCAP...06..002B,2013PhRvL.110b1301B}, which can be partially corrected if a lensing map is known. There was recently an interesting debate on the bias potentially introduced by second-order corrections: based on the calculations of Refs.~\cite{2014CQGra..31t2001U,2014CQGra..31t5001U} (see also Refs.~\cite{2005PhRvD..71f3537B,2012JCAP...11..045B,Andrianomena:2014sya} for earlier results), Ref.~\cite{2014JCAP...11..036C} suggested that second-order lensing could significantly affect the standard interpretation of the cosmic microwave background (CMB) observations. Nevertheless, this statement turned out to be inaccurate, due to confusions between several averaging schemes for the observable quantities at stake~\cite{2005ApJ...632..718K,2015arXiv150308506K,2015JCAP...07..040B,2015JCAP...06..050B}.

This problem of determining the effect of inhomogeneities on light propagation can also been tackled in a nonperturbative way, e.g. by relying on toy models. The most common examples are Swiss-cheese models~\cite{SW1,SW2}, where inhomogeneities are introduced within a background FL spacetime by inserting spherical patches of another exact solution of Einstein's equation. Recent analyses generally exploit the Lema\^{i}tre-Tolman-Bondi (LTB)~\cite{Marra:2007pm,Brouzakis:2007zi,2007JCAP...02..013B,Biswas:2007gi,2008PhRvD..77b3003M,Clifton:2009nv,Szybka:2010ky,Vanderveld:2008vi,Valkenburg:2009iw,2011JCAP...02..025B,Flanagan:2011tr,Flanagan:2012yv,2013PhRvL.110b1302B,2013JCAP...12..051L,2015arXiv150706590L} or Szekeres~\cite{2009GReGr..41.1737B,2010PhRvD..82j3510B,2014PhRvD..90l3536P,2014JCAP...03..040T} geometries as interior solutions, which aim at describing large-scale inhomogeneities such as superclusters or cosmic voids (see also Refs.~\cite{Adamek:2014qja,Bolejko:2012ue}). Observations have also been connected to the cosmic coarse-graining and backreaction issues in the series of works~\cite{1988A&A...206..190L,1998ApJ...497...28L,1998astro.ph..1122L,Rasanen:2008be,Rasanen:2009uw,Rasanen:2011bm,2011JCAP...07..008G,2012JCAP...04..036B,2014JCAP...10..073B}.

All the above-mentioned approaches have in common that they describe matter in the Universe as a fluid. However, when it comes to narrow beams, such as those involved in supernova (SN) observations, this approximation should no longer hold.\footnote{The typical physical size of a supernova explosion is on the order of a hundred astronomical units, which fixes the typical maximum cross-sectional diameter of the associated light beam. On such scales, the distribution of matter in the Universe cannot be considered smooth.} The applicability of the perturbation theory in this regime, in particular, has been questioned in Ref.~\cite{Clarkson:2011br}. This specific issue of how the \emph{clumpiness} of the Universe affects the interpretation of cosmological observables was first raised by Zel'dovich~\cite{Zeldo64} and Feynman~\cite{FeynmanColloquium}. The basic underlying idea is that in a clumpy medium, light mostly propagates through vacuum, and therefore experiences an underdense Universe. This stimulated a corpus of seminal articles~\cite{DashevskiiZeldovich1965,1965AZh....42..863D,1966SvA.....9..671D,1966RSPSA.294..195B,1967ApJ...150..737G,1967ApJ...147...61G,1976ApJ...208L...1W}, including the first analyses based on a Swiss-cheese model with Schwarzschild vacuoles~\cite{1969ApJ...155...89K,DR72,1973ApJ...180L..31D,1973PhDT........17D,1974ApJ...189..167D,1975ApJ...196..671R}. Contrary to LTB or Szekeres holes, the latter aim at modelling relatively small gravitationally bound structures, such as individual galaxies or stars. The analysis of light propagation in such models resulted in the so-called Dyer-Roeder approximation---that we shall rather call the Kantowski-Dyer-Roeder (KDR) approximation in this article, the name of Kantowski being unfairly omitted in the literature. Its correspondence with Swiss-cheese models has been carefully rederived and numerically checked in Ref.~\cite{Fleury:2014gha}, although its mathematical consistency was questioned in Refs.~\cite{Clarkson:2011br,Rasanen:2008be}. Analyses based on other models than Swiss cheeses, albeit physically similar in the sense that they also describe universes made of point masses, have been proposed in Refs.~\cite{Holz:1997ic,Okamura:2009zf,Bruneton:2012ru,Clifton:2009jw,2009JCAP...10..026C,2012PhRvD..85b3502C}. When applied to the interpretation of SN data, these various approaches generically do find a bias in the measurement of the cosmological parameters, on the order of a few to more than ten percent~\cite{1998ApJ...507..483K,Cooray:2005yr,Sarkar:2007sp,Fleury:2013uqa,Busti:2013sca}. It has been shown in Ref.~\cite{Fleury:2013uqa} that such an effect improves the agreement between SN and CMB observations regarding the measurement of $\Omega\e{m0}$.

While the KDR approximation may capture the main effects of the Universe's clumpiness on the average distance-redshift relation, it does not tell anything about its dispersion, and a fortiori about its higher-order moments. Model-based approaches do not in principle suffer from this weakness, but in all the works cited above, extracting e.g. the probability density function (PDF) of the observed angular distance at a fixed redshift requires numerical simulations which, because of their computational cost, lack of flexibility. A practical solution was proposed with the sGL method of Kainulainen and Marra~\cite{Kainulainen:2009dw,Kainulainen:2010at,2011PhRvD..84f3004K}, in which weak-lensing simulations have been maximally optimised so that generating $10^5$ mock observations only takes a few seconds. This method has been applied to forecast to which extent future SN observation campaigns, e.g. with the Large Synoptic Survey Telescope (LSST), would be able to constrain cosmological parameters from the moments of the distribution of SN magnitudes~\cite{2013PhRvD..88f3004M,2014PhRvD..89b3009Q,2015MNRAS.449.2845A,Amendola:2010ub}.
  
The goal of the present work is to propose an analytical and a priori non-perturbative framework for determining the statistical impact of small-scale structures on light propagation. Possible applications are the analysis of the bias and dispersion induced by these structures on cosmological observables, non only for distances measurements but also, e.g., cosmic shear. The main idea is that, on very small scales, the matter density field (i.e. the source of lensing) can be treated a \emph{white noise}, giving to lensing a diffusive behaviour. The equations of geometric optics in curved spacetime then take the form of generalised Langevin equations, which come with the whole machinery of statistical physics. Indeed, similar approaches have been exploited in other domains of physics~\cite{2013arXiv1305.4503N,1965PhRv..139..104H}, e.g., for describing the secular evolution of the Solar system. This systematic treatment of lensing as a stochastic process allows us to derive Fokker-Planck-Kolmogorov (FPK) equations for the PDF of the lensing observables, such as the angular distance, on which we will particularly focus in this article. 

The benefits of this new approach are multiple. Its analytical character potentially provides a better physical understanding of small-scale lensing, together with avoiding to rely on heavy ray-tracing simulations. It must be considered complementary to cosmic lensing due to the large-scale structure, with which it is planned to be merged in the future, in order to design a consistent multiscale description of lensing. Similarly to Refs.~\cite{2013PhRvD..88f3004M,2014PhRvD..89b3009Q,2015MNRAS.449.2845A,Amendola:2010ub}, we have in mind applications to a better characterisation of the matter distribution within the Universe. These various applications lie beyond the scope of the present article, which however proposes, as starters: (i) an extension of the KDR approximation, and (ii) an analytical calculation of the variance of the angular distance in an Einstein-Straus Swiss-cheese model.

The article is organised as follows. Section~\ref{sec:beams_two_formalisms} provides a theoretical lensing toolkit, which contains all the necessary material exploited in the remainder of the article, in particular the Jacobi matrix and the optical scalars. Sections~\ref{sec:Langevin} and~\ref{sec:FPK} are the heart of our approach: the former presents our fundamental hypotheses; the latter derives the FPK equations governing the PDF of the Jacobi matrix and of the optical scalars. Section~\ref{sec:general_results} deduces general analytical results from the FPK equations, in particular regarding the first two moments of the PDF of the angular distance. In order to test our formalism, we apply it to a Swiss-cheese model, and confront the associated predictions to numerical ray-tracing results in Section~\ref{sec:application_SC}. Section~\ref{sec:numerics} is finally devoted to a second check of our calculations, based on the numerical integration of the Langevin equation using the stochastic Euler method. It sheds some light of the connection between the accuracy of our predictions and the Gaussianity of the sources of lensing.

\section{Propagation of narrow light beams: two complementary formalisms}\label{sec:beams_two_formalisms}

Consider a narrow light beam, that is an infinitesimal bundle of null geodesics, converging at an observation event $O$. Among the geodesics of the bundle, we arbitrarily pick a reference ray~$\bar{x}^\mu(v)$, where $v$ is an affine parameter along the ray. The associated tangent vector $k^\mu\define \dd x^\mu / \dd v$ represents the wave four-vector of the light beam. If we choose $\vect{k}$ as \emph{past oriented} (so $v$ increases from $O$ to the source), then the (cyclic) frequency measured by an observer crossing the beam with four-velocity $\vect{u}$ is $\omega\define u^\mu k_\mu$. In this article, we set by convention~$v=0$ at $O$, and normalise all frequencies with respect to the observed one~$\omega_\obs\define (u^\mu k_\mu)|_O=1$.

The behaviour of any ray~$x^\mu(v)$ of the beam, relative to $\bar{x}^\mu(v)$, is characterised by its connecting vector $\xi^\mu\define x^\mu-\bar{x}^\mu$. If an observer at $\bar{x}^\mu(v)$ projects the beam on a screen, spanned by the Sachs basis (see Appendix~\ref{app:geometric_optics}), then the relative position of the two light spots associated with $\bar{x}^\mu$ and $x^\mu$ is a Euclidean two-dimensional vector $(\xi^A)_{A=1,2}$.

\subsection{Jacobi matrix}

The first standard tool for describing the effects of gravitational lensing is the Jacobi matrix, whose evolution with light propagation is a \emph{second-order linear differential equation}.

\subsubsection{Definition}

The Jacobi matrix is a $2\times 2$ matrix $\vect{\jacobi}=[\jacobi_{AB}]$ which relates the physical separation~$\xi^A$ (in screen space) between two rays with their angular separation~$\dot{\xi}^B(0)$---a dot denotes a derivative with respect to $v$---on the observer's celestial sphere, according to
\begin{equation}\label{eq:def_Jacobi}
\xi^A(v) = \jacobi\indices{^A_B}(v) \, \dot{\xi}^B(0) .
\end{equation}
The determinant of $\vect{\jacobi}$ thus represents the ratio between the beam's cross-sectional area~$A(v)=\dd^2 \xi(v)$ at $v$ with its observed angular aperture~$\Omega_\obs=\dd^2 \dot{\xi}$. When evaluated at the source event ($v=v_\source$), we recognise the definition of the (squared) angular diameter distance between the source and the observer
\begin{equation}
\det \vect{\jacobi}(v_\source)  = \frac{A_\source}{\Omega_\obs} \define D\e{A}^2.
\label{eq:Jacobi_DA}
\end{equation}
We recall that, if the number of photons is conserved during their travel from the source to the observer, then the angular diameter distance~$D\e{A}$ is related to the luminosity distance---used e.g. in the Hubble diagram of SNe---by the distance duality relation
\begin{equation}
D\e{L} = (1+z)^2 D\e{A}, 
\end{equation}
which involves the redshift~$z=(\omega_\source-\omega_\obs)/\omega_\obs$ between the emitted and observed frequencies.

The other three degrees of freedom of~$\vect{\jacobi}$ encode the deformations of the light beam, i.e. the deformations between the intrinsic source's shape and the observed image. This information is conveniently extracted from $\vect{\jacobi}$ by the decomposition given in Appendix~\ref{app:geometric_optics}.

\subsubsection{Evolution: the Jacobi matrix equation}

Because $\vect{\jacobi}$ describes the relative behaviour of two neighbouring light rays, its evolution with light propagation (i.e. with $v$) is inherited from the geodesic deviation equation; it results into the following second-order linear Jacobi matrix equation~\cite{2004LRR.....7....9P}
\begin{equation}\label{eq.jacobieq0}
 \ddot{\vect{\jacobi}}={\vect{\optical}}(v){\vect{\jacobi}}(v)
\end{equation}
where $\optical_{AB}\define R_{\mu\nu\rho\sigma} s_A^\mu k^\nu k^\rho s_B^\sigma$ is called the optical tidal matrix, and $(s_A^\mu)_{A=1,2}$ denotes the Sachs basis. The optical tidal matrix is symmetric due to the symmetries of the Riemann tensor~$R_{\mu\nu\rho\sigma}$.
The decomposition of the latter into a Ricci (trace) part and a Weyl (trace-free) part implies, for the optical tidal matrix,
\begin{equation}\label{eq:decomposition_optical_matrix}
\vect{\optical} = \Ricfoc \, \vect{1}_2 + \vect{\Weylmat},
\end{equation}
$\vect{1}_2$ standing for the $2\times 2$ unity matrix, while
\begin{align}
\Ricfoc 
&\define -\frac{1}{2} R_{\mu\nu} k^\mu k^\nu \\
\Weylmat_{AB}
&\define C_{\mu\nu\rho\sigma} s_A^\mu k^\nu  k^\rho s_B^\sigma,
\end{align}
where $R_{\mu\nu}$ and $C_{\mu\nu\rho\sigma}$ denote respectively the Ricci and Weyl tensors. It is straightforward to check that $\vect{\Weylmat}$ is trace free, and can thus be written as
\begin{equation}
\vect{\Weylmat}
=
\begin{pmatrix}
- \Weylfoc_1 & \Weylfoc_2 \\
 \Weylfoc_2 & \Weylfoc_1
\end{pmatrix},
\qquad \text{with} \quad
\Weylfoc_1+\i\Weylfoc_2
\define \Weylfoc
\define -\frac{1}{2} C_{\mu\nu\rho\sigma}
(s_1^{\mu} - {\rm i} s_2^{\mu}) k^\nu k^\rho (s_1^{\sigma} - {\rm i} s_2^{\sigma})
\end{equation}
The Ricci term, on the one hand, is directly related to the local energy-momentum density via the Einstein equation, $\Ricfoc=-4\pi G T_{\mu\nu}k ^\mu k^\nu \leq 0$ (under the null energy condition); it translates the isotropic focusing effect caused by smooth matter enclosed by the light beam. The Weyl term, on the other hand, essentially encodes tidal distortion effects, due to matter outside the beam, which tends to shear and rotate it.


The initial conditions ($v=0$) for Eq.~\eqref{eq.jacobieq0} are by definition [see Eq.~\eqref{eq:def_Jacobi}]
\begin{align}
\vect{\jacobi}(0) &= \zero_2
\label{eq:initial_Jacobi_1}\\
\dot{\vect{\jacobi}}(0) &= \identity_2,
\label{eq:initial_Jacobi_2}
\end{align}
so that, near the observer ($v\rightarrow 0$), the Jacobi matrix admits the expansion
\begin{equation}\label{eq:expansion_Jacobi_near_obs}
\vect{\jacobi}(v) = v \, \identity_2 + \frac{v^3}{3!} \, \vect{\optical}_\obs + \mathcal{O}(v^4) .
\end{equation}
It also implies, using that for any matrix~$\vect{M}$, $\det(1+\eps\vect{M})=1+\eps\,\tr\vect{M}+\mathcal{O}(\eps^2)$,
\begin{equation}\label{eq:expansion_DA_near_obs}
D\e{A}(v) = v + \frac{v^3}{3!} \, \Ricfoc_\obs + \mathcal{O}(v^4).
\end{equation}
%

\subsection{Optical scalars}

A standard alternative to the Jacobi matrix consists in a set of optical scalars, describing the deformation rate of the beam rather than net transformations. The resulting light propagation equations (Sachs equations) are a set of \emph{first-order nonlinear equations}.

\subsubsection{Definition}

The deformation rate of the light beam is naturally defined by a logarithmic derivative of the Jacobi matrix, namely through
\begin{equation}\label{eq.defS}
\vect{\deformation}\define \dot{\vect{\jacobi}} \vect{\jacobi}^{-1}.
\end{equation}
This deformation rate matrix can be shown to be symmetric, because of the symmetry of $\vect{\optical}$, and is thus decomposed as
\begin{equation}
\vect{\deformation}
=
\begin{pmatrix}
\theta & 0 \\
0 & \theta
\end{pmatrix}
+
\begin{pmatrix}
-\sigma_1 & \sigma_2 \\
\sigma_2 & \sigma_1
\end{pmatrix},
\label{eq:decomposition_deformation}
\end{equation}
where $\theta$ and $\sigma=\sigma_1+\i\sigma_2$ are the optical scalars, respectively called the expansion rate and the shear rate. The first one is directly related to the increase rate of the angular diameter distance, since $\dd(\ln\det{\vect{\jacobi}})/\dd v=\tr\vect{\deformation}$, i.e.
\begin{equation}\label{eq:relation_theta_DA}
\theta = \frac{\dot{D}\e{A}}{D\e{A}}.
\end{equation}


\subsubsection{Evolution: the Sachs scalar equations}

Inserting the definition~\eqref{eq.defS} into Eq.~\eqref{eq.jacobieq0} yields the evolution equation for $\vect{\deformation}$,
\begin{equation}
\dot{\vect{\deformation}} + \vect{\deformation}^2 = \vect{\optical},
\label{eq:evolution_deformation}
\end{equation}
from which the Sachs scalar equations follow:
\begin{align}
\dot{\nul{\theta}} + \nul{\theta}^2 + \abs{\nul{\sigma}}^2 &= \Ricfoc 
\label{eq:evolution_expansion}\\
\dot{\nul{\sigma}} + 2\nul{\theta} \nul{\sigma} &= \Weylfoc.
\label{eq:evolution_shear}
\end{align}
Using that $\theta=\dot{D}\e{A}/D\e{A}$, the above equation yields the so-called \emph{focusing theorem}
\begin{equation}\label{eq:focusing_theorem}
\ddot{D}\e{A} = (\Ricfoc - \abs{\sigma}^2) D\e{A},
\end{equation}
where we see that, while Ricci lensing has a direct focusing effect which tends to reduce $D\e{A}$, Weyl lensing has a similar but indirect effect, via the shear rate.

The initial conditions for the optical scalars are nontrivial, because $\vect{\jacobi}$ vanishes for $v=0$, which implies that $\vect{\deformation}$ must have a pole at the observation event. Precisely, the initial behaviour~\eqref{eq:expansion_Jacobi_near_obs} of the Jacobi matrix yields
\begin{equation}\label{eq:initial_deformation}
\vect{\deformation}(v) 
= \pac{ \identity_2 + \mathcal{O}(v^2) } \pac{ v\,\identity_2 + \mathcal{O}(v^3) }^{-1}
= v^{-1}\identity_2 + \mathcal{O}(v),
\end{equation}
and we conclude that the initial conditions ($v\rightarrow 0$) for the optical scalars are
\begin{align}
\nul{\theta}(v) &= \frac{1}{v} + \mathcal{O}(v), \\
\nul{\sigma}(v) &= \mathcal{O}(v).
\end{align}
Hence only the expansion rate has a pole at $v=0$, while the shear rate is regular.

\section{Small-scale lensing as a diffusion process}\label{sec:Langevin}

We now focus on the specific issue of lensing caused by the small-scale inhomogeneity of the Universe, i.e, down to scales where the matter distribution experienced by the light beam cannot be considered a continuous medium, but rather by a multitude of mass clumps that all slightly distort it. This situation is analogous to the Brownian motion of a particle suspended in water, where a macroscopic---continuous-medium---description of the liquid is no longer sufficient, and must be replaced by a semi-microscopic approach in order to account for the collisions between the particle and water molecules.

The approach developed in the present article is based on this analogy. Just like in the standard treatment of the Brownian motion, where particle-molecule collisions are modelled by a stochastic force, we propose to introduce stochastic terms in the lensing scalars $\Ricfoc$, $\Weylfoc$. The equations governing light propagation will thus take the form of Langevin equations.

\subsection{Fundamental hypotheses}\label{sec:hypotheses} 

We split the Ricci and Weyl lensing scalars experienced by the light beam into a deterministic part representing their average, slowly varying behaviour, and a stochastic part modelling their rapid fluctuations:
\begin{align}
\Ricfoc &= \ev{\Ricfoc} + \delta\Ricfoc , \\
\Weylfoc &= \ev{\Weylfoc} + \delta\Weylfoc,
\end{align}
where $\ev{\ldots}$ is an ensemble average, and $\ev{\delta\Ricfoc}=\ev{\delta\Weylfoc}=0$. All these quantities are in principle functions of the affine parameter. Note that, despite the notation, \emph{$\delta\Ricfoc$ and $\delta\Weylfoc$ are not necessarily small with respect to $\ev{\Ricfoc}$ and $\ev{\Weylfoc}$}, they are not dealt with as perturbations. The deterministic components can be thought of as the optical properties of an average universe, in the sense e.g. of Ref.~\cite{2014JCAP...10..073B}---a notion which may not coincide with a spatial average, or with a FL model.

We now make the following hypotheses:
\begin{description}
\item[Azimuthal symmetry about the beam.] We suppose that the Universe is statistically homogeneous and isotropic, which implies statistical symmetry with respect to rotations about any light beam. This motivates us to assume that the direction along which a beam is sheared is \emph{independent} from the shear amplitude. It is also independent from Ricci focusing. In other words, decomposing the Weyl lensing scalar as $\Weylfoc=|\Weylfoc|\ex{-2\i\beta}$, we assume that $\beta$ is statistically independent from $|\Weylfoc|$ and $\Ricfoc$. However, we emphasize that $\abs{\Weylfoc}$ is \emph{not} independent from $\Ricfoc$.
\item[Statistical isotropy.] We suppose that the Universe has no preferred (spatial) direction, which implies that $\beta$ must be uniformly distributed in $[0,\pi]$. As a consequence,
\begin{equation}
\ev{\Weylfoc} = \ev{|\Weylfoc|} \ev{ \ex{-2\i\beta} }  = 0,
\end{equation}
where we have also used our first hypothesis. We can thus omit the $\delta$ in the stochastic part of $\Weylfoc$. Furthermore, for any $v,w$
\begin{align}
\ev{\delta\Ricfoc(v) \Weylfoc(w)} &= \ev{\delta\Ricfoc(v) |\Weylfoc(w)|}\ev{\ex{-2\i\beta(w)}} = 0,\\
\ev{\Weylfoc_1(v) \Weylfoc_2(v)} &= \frac{1}{2}\ev{|\Weylfoc(v)|^2}\ev{\sin 4\beta(v)} = 0.
\end{align}
\item[White noises.] Because they model rapidly fluctuating functions, the coherence scale of $\delta\Ricfoc$ and $\Weylfoc$ is much smaller than the typical evolution scale of the Jacobi matrix, of the optical scalars, and than the typical distance between the source and the observer. Therefore, they can be considered \emph{white noises}, i.e. \emph{$\delta$-correlated Gaussian random processes}\footnote{
A random process~$t\mapsto X(t)$ is Gaussian if any of its finite-dimensional probability distributions is a multivariate Gaussian,
\begin{equation}
p_{t_1,\ldots t_n}(x_1,\ldots x_n) \propto \exp \pa{ -\frac{1}{2} \sum_{i,j=1}^n x_i C^{-1}_{ij} x_j },
\end{equation}
where $C_{ij}=C(t_i,t_j)\define\ev{ X(t_i) X(t_j) }$ is the covariance of the process, and $C^{-1}$ denotes its inverse. A white noise corresponds to the limit where $C(t_i,t_j)\propto \delta(t_i,t_j)$. Hence, for a white noise, $X(t_1)$ and $X(t_2\not= t_1)$ are independent.
}, with
\begin{align}
\ev{\delta \Ricfoc(v) \delta \Ricfoc(w)} &= C_\Ricfoc(v) \delta(v-w) 
\label{eq:def_covariance_Ricci}\\
\ev{\Weylfoc_A(v) \Weylfoc_B(w)} &= C_\Weylfoc(v) \delta_{AB} \delta(v-w),
\label{eq:def_covariance_Weyl}
\end{align}
where the $\delta_{AB}$ in Eq.~\eqref{eq:def_covariance_Weyl} comes from statistical isotropy. The functions $C_\Ricfoc, C_\Weylfoc$ shall be called the \emph{covariance amplitudes} of Ricci and Weyl lensing. Gaussianity, which is motivated by the central limit theorem, ensures that $\delta\Ricfoc(v)$ [resp. $\Weylfoc(v)$] and $\delta\Ricfoc(w\not=v)$ [resp. $\Weylfoc(w\not=v)$] are not only uncorrelated, but also independent.
\end{description}

Physically speaking, the covariance amplitude $C_X$ of the white noise~$X(t)$ modelling a physical process~$X\e{phys}(t)$ must be understood as $C_X\sim (\delta X\e{phys})^2 \Delta t\e{coh}$, where $\delta X\e{phys}$ is the typical fluctuation amplitude of $X\e{phys}$, while $\Delta t\e{coh}$ is the scale on which it remains coherent. For classical Brownian motion, this scale corresponds to the duration of a typical particle-molecule collision; in gravitational lensing, it will represent the typical extension of a gas cloud/dark matter halo (Ricci lensing), or the affine-parameter length over which the beam undergoes the tidal influence of a given deflector (Weyl lensing).

In principle, the deterministic components $\ev{\Ricfoc}$ and $\ev{\Weylfoc}$ could also allow for the large-scale structure of the Universe (cosmic voids, walls, and filaments). For simplicity, we do not consider this possibility in the present paper, and focus our attention on the rapidly fluctuating terms. It will be convenient, in the following, to gather them into a 3-dimensional \emph{noise vector}~$\vect{N}$ such that
\begin{equation}\label{eq.defnoise}
\transpose{\vect{N}}\define(\delta\Ricfoc,\Weylfoc_1,\Weylfoc_2).
\end{equation}
We also introduce the \emph{diffusion matrix}~$\vect{\diffusion}$ of $\vect{N}$, defined by\footnote{Equivalently, the diffusion matrix can be defined from the increments of the Brownian motion~$\vect{B}$ associated with $\vect{N}$, i.e. such that $\dd\vect{B}=\vect{N}\dd v$. Between $v_1$ and $v_2$, the increment of $\vect{B}$ is $\Delta\vect{B}\define \vect{B}(v_2)-\vect{B}(v_1)$, and its variance reads $\ev{\Delta\vect{B}\Delta\transpose{\vect{B}}}=\vect{\diffusion}\Delta v$, with $\Delta v\define v_2-v_1$.} $\ev{\vect{N}(v)\transpose{\vect{N}}(w)}=\vect{\diffusion}(v)\delta(v-w)$, which here reads
\begin{equation}
\label{eq:Diff_Matrix}
\vect{\diffusion}=\mathrm{diag}(C_\Ricfoc,C_\Weylfoc,C_\Weylfoc).
\end{equation}

\subsection{Langevin equation for the Jacobi matrix}\label{sec:Langevin_Jacobi}

The Jacobi matrix equation~\eqref{eq.jacobieq0} reads
\begin{equation}
\ddot{\vect{\jacobi}}= \ev{\Ricfoc}\vect{\jacobi} + \pa{\delta \Ricfoc + \vect{{\Weylmat}}} \vect{\jacobi},
\end{equation}
where we have separated the deterministic and stochastic terms on the right-hand side. It is analogous to a system of coupled harmonic oscillators with fluctuating stiffness. Some further insights on this dynamical system can be obtained thanks to a Hamiltonian formulation
\begin{equation}
\begin{system}
\dot{\jacobi}_{AB} &= {\cal P}_{AB} = \frac{\partial H}{\partial{\cal P}_{AB}} \\
\dot{\cal P}_{AB} &= -\frac{\partial V\e{Jac}}{{\cal D}_{AB}}  = -\frac{\partial H}{\partial{\jacobi}_{AB}}  + \mathcal{N}_{AB}(v)
\end{system},
\end{equation}
with
\begin{equation}
H \equiv \frac{1}{2} \tr\pa{ \transpose{\vect{\mathcal{P}}} \vect{\mathcal{P}} 
										-\ev{\Ricfoc}\transpose{\vect{\jacobi}}\vect{\jacobi} },
\qquad
\vect{\mathcal{N}} \define \pa{\delta \Ricfoc \vect{1}_2 + \vect{{\Weylmat}}} \vect{\jacobi},
\end{equation}
and where the Hamiltonian~$H$ encodes only the non-stochastic part of the process. Such a dynamics is very similar to the integrable systems with stochastic perturbations discussed e.g. in Ref.~~\cite{2013arXiv1305.4503N}, except that (i) due to the explicit $v$-dependence of $H$, through $\ev{\Ricfoc}$, the unperturbed system is not fully integrable; and (ii) the stochastic term~$\vect{\mathcal{N}}$ contains the variable~$\vect{\jacobi}$: the noise is \emph{multiplicative}. This analogy with dynamical systems in statistical mechanics also provides a nice interpretation of the deformation rate matrix~$\vect{\deformation}$: as a Ricatti variable associated with $\vect{\jacobi}$, it defines the so-called Kolmogorov-Sinai entropy of the random process, $h\e{KS}=\tr(\vect{\deformation})$.

Let us now put the Jacobi matrix equation in the form of a first-order \emph{Langevin} equation, which will be useful for deriving the associated Fokker-Planck-Kolmogorov equations in Sec.~\ref{sec:FPK}. For that purpose, we first need to vectorise the Jacobi matrix as
\begin{equation}
\vect{D} \define (D_\alpha)_{\alpha\in\{1\ldots 4\}}
\qquad
\text{with}
\qquad
\jacobi_{AB} = D_{2(A-1)+B};
\end{equation}
in other words, we represent the couples of matrix indices $(AB)$ by one single index~$\alpha$, so that $1=(11)$, $2=(12)$, $3=(21)$, $4=(22)$. We then construct an 8-dimensional vector $\transpose{\vect{J}}\define(\vect{D},\vect{\dot{D}})$, whose dynamics is described by the Langevin equation
\begin{equation}
\ddf{\vect{J}}{v} = \vect{M}(v) \vect{J}(v) + \vect{L}\e{Jac}(\vect{J}) \vect{N}(v).
\label{eq:Langevin_matrix}
\end{equation}
where the \emph{drift} matrix is
\begin{equation}
\vect{M}
\define
\begin{bmatrix}
\vect{0}_4 & \vect{1}_4 \\
\ev{\Ricfoc}\vect{1}_4 & \vect{0}_4
\end{bmatrix},
\end{equation}
and the \emph{noise-mixing} matrix reads
\begin{equation}\label{eq:L_Jac}
\vect{L}\e{Jac}
\define
\begin{bmatrix}
 & \zero_{4\times 3} &  \\
D_1 & -D_1 & D_3 \\
D_2 & -D_2 & D_4 \\
D_3 & D_3 & D_1 \\
D_4 & D_4 & D_2
\end{bmatrix}
=
\begin{bmatrix}
 & \zero_{4\times 3} &  \\
\jacobi_{11} & -\jacobi_{11} & \jacobi_{21} \\
\jacobi_{12} & -\jacobi_{12} & \jacobi_{22} \\
\jacobi_{21} & \jacobi_{21} & \jacobi_{11} \\
\jacobi_{22} & \jacobi_{22} & \jacobi_{12}
\end{bmatrix}.
\end{equation}
Equation~\eqref{eq:Langevin_matrix} is linear, with a multiplicative noise.

\subsection{Langevin equation for the optical scalars}\label{subsec:scalar_formalism}

A similar procedure can be achieved for the optical scalars. The Sachs equations~(\ref{eq:evolution_expansion}-\ref{eq:evolution_shear}), together with the relation~\eqref{eq:relation_theta_DA} between the angular distance and the beam's expansion rate, form the system
\begin{align}
\dot{D}\e{A} &= \nul{\theta} D\e{A}, \\
\dot{\nul{\theta}} &= -\nul{\theta}^2 - \abs{\nul{\sigma}}^2 + \ev{\Ricfoc} + \delta \Ricfoc, \\
\dot{\nul{\sigma}} &= -2\nul{\theta}\nul{\sigma} + \Weylfoc,
\end{align}
which, defining the 4-dimensional vector $\transpose{\vect{S}}\define(D\e{A}, \nul{\theta},\nul{\sigma}_1,\nul{\sigma}_2)$, becomes the \emph{Sachs-Langevin} equation
\begin{equation}
\ddf{\vect{S}}{v} = \vect{F}(v,\vect{S}) + \vect{L}\e{scal}\vect{N}(v),
\label{eq:Langevin_scalar}
\end{equation}
where the drift term reads $\transpose{\vect{F}}\define(\nul{\theta}D\e{A},-\nul{\theta}^2-\abs{\nul{\sigma}}^2+\ev{\Ricfoc},-2\nul{\theta}\nul{\sigma}_1,-2\nul{\theta}\nul{\sigma}_2)$, while the noise mixing matrix is
\begin{equation}
\vect{L}\e{scal}
\define
\begin{bmatrix}
0 & 0 & 0\\
1 & 0 & 0\\
0 & 1 & 0\\
0 & 0 & 1
\end{bmatrix}.
\end{equation}
Contrary to Eq.~\eqref{eq:Langevin_matrix}, Eq.~\eqref{eq:Langevin_scalar} has a nonlinear drift term (which reflects the nonlinearity of the Sachs scalar equations), but its noise is additive, in the sense that the stochastic term $\vect{L}\e{scal}\vect{N}$ is independent of the variable~$\vect{S}$.

\section{The lensing Fokker-Planck-Kolmogorov equations}\label{sec:FPK}

The presence of stochastic terms in the optical equations gives a diffusive behaviour to the lensing observables, which can be quantified by their PDFs. When a dynamical system is ruled by a Langevin equation, its PDF in phase space satisfies a partial differential equation called the Fokker-Planck-Kolmogorov (FPK) equation. In \S~\ref{sec:from_Langevin_to_FPK}, we recall the general procedure to derive the FPK equation associated with a Langevin equation; we then apply it to the Jacobi matrix (\S~\ref{sec:FPK_matrix}) and to the optical scalars (\S~\ref{sec:FPK_scalar}).

\subsection{From Langevin to Fokker-Planck-Kolmogorov}\label{sec:from_Langevin_to_FPK}

Consider the following general Langevin equation governing the evolution of a $n$-dimensional random process~$t\mapsto\vect{X}(t)$,
\begin{equation}\label{eq:Langevin_general}
\frac{ {\rm d}\vect{X}}{\dd t} = \vect{f}(\vect{X},t)  +   \vect{L}(\vect{X},t)\vect{N}(t),
\end{equation}
where the $n$-dimensional vector~$\vect{f}$ and the $n\times n$ matrix~$\vect{L}$ are deterministic, while $\vect{N}$ is a white noise. One can easily see that both our Langevin equations~\eqref{eq:Langevin_matrix} and \eqref{eq:Langevin_scalar} have this form, the affine parameter playing the role of time~$t$, and the random process being either $\vect{J}$ or $\vect{S}$. The mathematical difficulty of Eq.~\eqref{eq:Langevin_general} is that it cannot be treated with the ordinary theory of differential equations, because $\vect{N}(t)$ is discontinuous everywhere. In general, the solution of Eq.~\eqref{eq:Langevin_general} is not unique, even for a given realization of $\vect{N}$.

A standard approach~\cite{feller71,arnold74,ks1991,ls2001,oksendal2003,1989fpem.book.....R} consists in introducing the It\=o calculus, the main properties of which we summarise below. One can formally integrate Eq.~(\ref{eq:Langevin_general}) as
\begin{equation}\label{eq.ex2}
\vect{X}(t) - \vect{X} (t_0) = \int_{t_0}^t \vect{f}(\vect{X}, t)\;\dd t +  \int_{t_0}^t\vect{L}(\vect{X},t)\;\vect{N}(t)\dd t,
\end{equation}
where the second integral requires particular attention, because the Riemann or Lebesgue definitions cannot apply, due to the unboundedness and discontinuity of $\vect{N}$. First, it must be reformulated as a Stieltjes integral
\begin{equation}
\int_{t_0}^t\vect{L}(\vect{X},t)\;\dd \vect{B}
\label{eq:stochastic_integral}
\end{equation}
where $\vect{B}$ is a \emph{Brownian motion}, i.e. a stochastic process whose any increment $\Delta\vect{B}_k\equiv\vect{B}(t_{k+1})-\vect{B}(t_{k})$ is a zero mean Gaussian random variable with variance $\ev{\Delta\vect{B}_k\transpose{\Delta\vect{B}}_k}=\vect{\diffusion}(t_k,t_{k+1}) \Delta t_k$. $\vect{\diffusion}$ is called the \emph{diffusion matrix} of $\vect{B}$.
The white noise $\vect{N}$ is thus considered a formal derivative of the Brownian motion~$\vect{B}$, i.e. $\dd\vect{B}=\vect{N}\dd t$. One possible definition for the integral~\eqref{eq:stochastic_integral} follows the so-called  It\=o stochastic prescription~\cite{kean69},
\begin{equation}
\int_{t_0}^t\vect{L}(\vect{X},t) \; \dd \vect{B} \define
\lim_{n\to\infty} \sum_{k=0}^{n-1} \vect{L}[\vect{X}(t_k),t_k] \left[\vect{B}(t_{k+1})-\vect{B}(t_{k})\right].
\end{equation}

This definition leads to some modifications with respect to ordinary differential calculus when $\vect{B}$ is involved. For example, it can be shown by calculating explicitly the It\=o integral of $B_i \dd B_j$ that $\dd(B_i B_j)=B_i\dd B_j+B_j\dd B_i + \diffusion_{ij}\dd t$,  which implies
\begin{equation}
\dd\vect{B} \, \transpose{\dd\vect{B}} = \vect{\diffusion} \, \dd t .
\label{eq:dB_squared}
\end{equation}
The above quantity is thus of order $1$ in $\dd t$, contrary to what we would naively expect by replacing $\dd\vect{B}$ by $\vect{N}\dd t$. Equation~\eqref{eq:dB_squared} is the most important rule of the It\=o calculus. As a consequence, the first-order Taylor expansion of any function~$\phi(t,\vect{X})$ must actually include second-order terms $\propto \dd X_i \, \dd X_j$, since
\begin{equation}
{\rm d}\vect{X} = \vect{f}(\vect{X},t)  {\rm d}t +   \vect{L}(\vect{X},t)\,\dd\vect{B},
\end{equation}
contains $\dd\vect{B}$. More precisely,
\begin{align}
\dd\phi &= \pd{\phi}{t} \dd t + \pd{\phi}{X_i} \dd X_i 
					+ \frac{1}{2} \frac{\partial^2\phi}{\partial X_i \partial X_j} \dd X_i \dd X_j \\
			&= \pa{ \pd{\phi}{t} + \frac{1}{2} \frac{\partial^2\phi}{\partial X_i \partial X_j} L_{ik} \diffusion_{kl} L_{jl} } \dd t 
					+ \pd{\phi}{X_i} \dd X_i,
\label{eq:Ito_formula}
\end{align}
which is known as the It\=o formula~\cite{feller71,arnold74,ks1991,ls2001,oksendal2003,kean69,1989fpem.book.....R} .

From the It\=o formula, one can deduce the Fokker-Planck-Kolmogorov (FPK) equation governing the PDF~$p(t;\vect{X})$ of the stochastic process~$\vect{X}(t)$. The derivation~\cite{arnold74,ls2001} relies on a trick which consists in inserting Eq.~\eqref{eq:Ito_formula} in the time derivative of the expectation value of an arbitrary function~$\phi(t,\vect{X})$,
\begin{equation}
\ev{\phi}(t) \define \int \phi(t,\vect{X}) \, p(t;\vect{X}) \; \dd^n\vect{X},
\end{equation}
which, after a few integration by parts, yields
\begin{equation}\label{eq:FPK_general}
\frac{\partial p(t;\vect{X})}{\partial t}
= -\frac{\partial}{\partial X_i} [f_i(\vect{X},t) p(t;\vect{X})]
	+\frac12 \frac{\partial^2}{\partial X_i \partial X_j} \left\lbrace\left[ \vect{L}(t;\vect{X}) \vect{Q}(t) \transpose{\vect{L}}(\vect{X},t) \right]_{ij} p(t;\vect{X})\right\rbrace.
\end{equation}
The first term on the right-hand side is a drift term, it drives the global displacement of the probability packet, while the second is a diffusion term, which tends to spread it. With this summary of textbook results~\cite{feller71,arnold74,ks1991,ls2001,oksendal2003,kean69,1989fpem.book.....R} we wish to emphasize that the derivation of the FPK equation requires the noise to be white, i.e. $\vect{N}=\dd\vect{B}/\dd t$ where $\vect{B}$ is a Brownian motion, so that the It\=o calculus can be applied. The hypotheses formulated in \S~\ref{sec:hypotheses} are therefore crucial for this formalism to be applicable.

\subsection{FPK equation for the Jacobi matrix}\label{sec:FPK_matrix}

Let us now derive the FPK equation governing the PDF of the Jacobi matrix. Applying the general formula~\eqref{eq:FPK_general} to the Langevin equation~\eqref{eq:Langevin_matrix} leads to the following equation for the PDF~$p(v;\vect{J})$,
\begin{equation}
\pd{p}{v} = -\pd{}{J_a} \pa{ M_{ab} J_b \, p} 
					+ \frac{1}{2} \frac{\partial}{\partial J_a \partial J_b} 
										\pac{ \pa{\vect{L}\e{Jac} \vect{\diffusion} \transpose{\vect{L}}\e{Jac}}_{ab} p },
\end{equation}
where the indices $a,b$ run from 1 to 8. Using the explicit expression~\eqref{eq:L_Jac} of $\vect{L}\e{Jac}$, we can write the matrix involved in the diffusion term as
\begin{equation}
\vect{L}\e{Jac}\vect{\diffusion}\transpose{\vect{L}}\e{Jac}
=
\begin{bmatrix}
\zero_4 & \zero_4 \\
\zero_4 & \vect{\Gamma}
\end{bmatrix},
\end{equation}
where the components of the $4\times 4$ symmetric matrix $\vect{\Gamma}$ are
\begin{align}
\Gamma_{11} &= (C_\Ricfoc + C_\Weylfoc) \jacobi_{11}^2 + C_\Weylfoc \jacobi_{21}^2\nonumber \\
\Gamma_{12} &= (C_\Ricfoc + C_\Weylfoc) \jacobi_{11}\jacobi_{12} + C_\Weylfoc \jacobi_{21}\jacobi_{22} = \Gamma_{21}\nonumber\\
\Gamma_{13} &= C_\Ricfoc \jacobi_{21}\jacobi_{11} = \Gamma_{31}\nonumber\\
\Gamma_{14} &= (C_\Ricfoc - C_\Weylfoc) \jacobi_{11}\jacobi_{22} + C_\Weylfoc \jacobi_{21}\jacobi_{12} = \Gamma_{41}\nonumber\\
\Gamma_{22} &= (C_\Ricfoc + C_\Weylfoc) \jacobi_{12}^2 + C_\Weylfoc \jacobi_{22}^2\nonumber\\
\Gamma_{23} &= (C_\Ricfoc - C_\Weylfoc) \jacobi_{12}\jacobi_{21} + C_\Weylfoc \jacobi_{11}\jacobi_{22} = \Gamma_{32}\nonumber\\
\Gamma_{24} &= C_\Ricfoc \jacobi_{12}\jacobi_{22} = \Gamma_{42}\nonumber\\
\Gamma_{33} &= (C_\Ricfoc + C_\Weylfoc) \jacobi_{21}^2 + C_\Weylfoc \jacobi_{11}^2\nonumber\\
\Gamma_{34} &= (C_\Ricfoc + C_\Weylfoc) \jacobi_{21}\jacobi_{22} + C_\Weylfoc \jacobi_{11}\jacobi_{12} = \Gamma_{43}\nonumber\\
\Gamma_{44} &= (C_\Ricfoc + C_\Weylfoc) \jacobi_{22}^2 + C_\Weylfoc \jacobi_{12}^2.
\end{align}
A few calculations and reorganizations yield the following explicit form of the FPK equation of $p(v;\vect{J})=p(v;\vect{\jacobi},\vect{\dot{\jacobi}})$,
\begin{empheq}[box=\fbox]{multline}
\pd{p}{v} = - \dot{\jacobi}_{AB} \pd{p}{\jacobi_{AB}}
					- \ev{\Ricfoc} \jacobi_{AB} \frac{\partial p}{\partial \dot{\jacobi}_{AB}} \\
				+ \frac{1}{2} \pac{ C_\Ricfoc\,\delta_{AE} \delta_{CF} + C_\Weylfoc (\delta_{AC} \delta_{EF} - \eps_{AC}\eps_{EF})} 
					 	\jacobi_{EB}\jacobi_{FD}\,\frac{\partial^2 p}{\partial\dot{\jacobi}_{AB}\partial\dot{\jacobi}_{CD}}
\label{eq:FPK_matrix}
\end{empheq}
where $\eps_{AB}$ is the two-dimensional antisymmetric matrix with $\eps_{12}=1$. Equation~\eqref{eq:FPK_matrix} can also be rewritten in an elegant formal way as
\begin{multline}
\pd{p}{v} = \Bigg\{ - \tr\pa{ \vect{\dot{\jacobi}}^T\pd{}{\vect{\jacobi}} }
							- \ev{\Ricfoc}\tr\pa{ \vect{\jacobi}^T\pd{}{\vect{\dot{\jacobi}}} }
			+ \frac{C_\Ricfoc}{2} \tr\pac{\pa{ \vect{\jacobi}^T\pd{}{\vect{\dot{\jacobi}}} }^2} \\
		+ \frac{C_\Weylfoc}{2} \pac{ \tr\pa{ \vect{\jacobi}^T\pd{}{\vect{\dot{\jacobi}}} } }^2
		- C_\Weylfoc \det\pa{ \vect{\jacobi}^T\pd{}{\vect{\dot{\jacobi}}} }
							\Bigg\} \, p
\label{eq:FPK_matrix_formal}
\end{multline}
which involves in particular the $2\times 2$ matrix differential operator
\begin{equation}
\pa{ \vect{\jacobi}^T\pd{}{\vect{\dot{\jacobi}}} }_{AB} \define \jacobi_{CA} \pd{}{\dot{\jacobi}_{CB}}.
\end{equation}
Finally, the boundary condition for Eq.~\eqref{eq:FPK_matrix} is deduced from the initial conditions~\eqref{eq:initial_Jacobi_1}, \eqref{eq:initial_Jacobi_2}, and reads
\begin{equation}
p(0;\vect{\jacobi},\dot{\vect{\jacobi}}) = \delta(\vect{\jacobi}) \delta(\dot{\vect{\jacobi}}-\identity_2).
\end{equation}

\subsection{FPK for the optical scalars}\label{sec:FPK_scalar}

Regarding optical scalars, starting from the Langevin equation~\eqref{eq:Langevin_scalar}, one can derive the following FPK equation for~$p(v;\vect{S})=p(v;D\e{A},\nul{\theta},\nul{\sigma}_1,\nul{\sigma}_2)$,
\begin{equation}\label{eq:FPK_scalar_0}
\pd{p}{v} = 
- \pd{F_\alpha p}{S_\alpha} 
+ \frac12 \frac{\partial^2}{\partial S_\alpha \partial S_\beta} 
				\pac{ \pa{ \vect{L}\e{scal}\vect{\diffusion}\transpose{\vect{L}}\e{scal} }_{\alpha\beta} p },
\end{equation}
where $\alpha,\beta$ run from $1$ to $4$, and where the diffusion term reads
\begin{equation}
\vect{L}\e{scal}\vect{\diffusion}\transpose{\vect{L}}\e{scal}
=
\begin{bmatrix}
0 & 0 & 0 & 0 \\
0 & C_\Ricfoc & 0 & 0 \\
0 & 0 & C_\Weylfoc & 0 \\
0 & 0 & 0 & C_\Weylfoc
\end{bmatrix}.
\end{equation}
It follows that Eq.~\eqref{eq:FPK_scalar_0} takes the explicit form
\begin{empheq}[box=\fbox]{multline}
\pd{p}{v} = -\nul{\theta} \pd{D\e{A}p}{D\e{A}} 
					+ \pd{}{\nul{\theta}}\pac{ \pa{\nul{\theta}^2 + \abs{\nul{\sigma}}^2 - \ev{\Ricfoc}} p }
					+ 2\nul{\theta} \pa{ \pd{\nul{\sigma}_1 p}{\nul{\sigma}_1} + \pd{\nul{\sigma}_2p}{\nul{\sigma}_2} } \\
					+ \frac{C_\Ricfoc}{2} \pd[2]{p}{\nul{\theta}}
					+ \frac{C_\Weylfoc}{2} \pa{ \pd[2]{p}{\nul{\sigma}_1} + \pd[2]{p}{\nul{\sigma}_2} }.
\label{eq:FPK_scalar}
\end{empheq}
The initial condition for $\theta$ being singular, it is not possible to write a boundary condition for Eq.~\eqref{eq:FPK_scalar} as we did for Eq.~\eqref{eq:FPK_matrix}.

\section{General analytical results}\label{sec:general_results}

Because it is a partial differential equation, the FPK equation is generally impossible to solve analytically, except in a few known special cases~\cite{1989fpem.book.....R}. Nevertheless, it can be used to derive evolution equations for the moments of the PDF, some of which are solvable. In this section, we derive some general analytical formulae on the moments of lensing observables. The results for the Jacobi matrix (\S~\ref{sec:moments_Jacobi}) and for the optical scalars (\S~\ref{sec:moments_scalars}) will turn out to be complementary, and used for deriving an evolution equation for the variance of the angular diameter distance in \S~\ref{sec:variance_angular_distance}.

\subsection{Moments of the Jacobi matrix distribution}\label{sec:moments_Jacobi}

The Jacobi matrix formalism has this considerable advantage on the optical scalar formalism that it enjoys a \emph{linear} Langevin equation. Despite the fact that its noise is multiplicative, this implies that all the moments of order-$n$ of the PDF of $\vect{\jacobi}$ satisfy a \emph{closed} system of differential equations. It is not the case when nonlinearities are present, in which case emerges a hierarchy of equations, where the evolution of the lower-order moments depends on moments of higher-order. 

\subsubsection{Order-one moments}

Let us start by deriving the evolution equations for the expectation values~$\ev{\vect{\jacobi}}$ and $\ev[1]{\vect{\dot{\jacobi}}}$. We proceed by multiplying the FPK equation~\eqref{eq:FPK_matrix} by $\jacobi_{IJ}$ (or $\dot{\jacobi}_{IJ}$) and then integrating it with respect to $\vect{\jacobi}$ and $\vect{\dot{\jacobi}}$. For $\jacobi_{IJ}$, this procedure yields
\begin{multline}
\ddf{}{v} \int \jacobi_{IJ} \, p \; \dd^4\vect{\jacobi} \,\dd^4\vect{\dot{\jacobi}}
=
- \int \jacobi_{IJ} \pd{\dot{\jacobi}_{AB} \, p}{\jacobi_{AB}} \; \dd^4\vect{\jacobi} \,\dd^4\vect{\dot{\jacobi}}
- \ev{\Ricfoc} \int \jacobi_{IJ} \frac{\partial  \jacobi_{AB} \, p}{\partial \dot{\jacobi}_{AB}} \; \dd^4\vect{\jacobi} \,\dd^4\vect{\dot{\jacobi}} \\
+ \frac{1}{2} \int \jacobi_{IJ} \frac{\partial^2}{\partial\dot{\jacobi}_{AB}\partial\dot{\jacobi}_{CD}}
					\paac{\pac{ C_\Ricfoc\,\delta_{AE} \delta_{CF} 
							+ C_\Weylfoc \pa{\delta_{AC} \delta_{EF} - \eps_{AC}\eps_{EF} }
							 		} \jacobi_{EB}\jacobi_{FD} \, p
							 } 
 \; \dd^4\vect{\jacobi} \,\dd^4\vect{\dot{\jacobi}}.
\label{eq:mean_Jacobi_intermediate}
\end{multline}
The left-hand side is clearly $\dd\ev{\jacobi_{IJ}}/\dd v$. On the right-hand side, the first term can be integrated by parts to give $\ev[1]{\dot{\jacobi}_{IJ}}$; the other two vanish since they can both be written as the integral of a derivative with respect to $\dot{\jacobi}_{AB}$. Equation~\eqref{eq:mean_Jacobi_intermediate} is thus simply
\begin{equation}
\ddf{\ev{\vect{\jacobi}}}{v} = \ev[1]{\vect{\dot{\jacobi}}}
\end{equation}
as one can intuitively expect.

The same method applied to $\dot{\jacobi}_{IJ}$ leads to
\begin{equation}
\ddf{\ev[1]{\vect{\dot{\jacobi}}}}{v} = \ev{\Ricfoc}\ev{\vect{\jacobi}},
\end{equation}
so that the expectation value of the Jacobi matrix reads
\begin{equation}
\ddf[2]{\ev{\vect{\jacobi}}}{v} = \ev{\Ricfoc}\ev{\vect{\jacobi}}.
\label{eq:evolution_ev_Jacobi}
\end{equation}
Note that this result could also have been obtained by directly averaging the Sachs-Langevin equation. However, this naive method would not work for higher-order moments, which is why we preferred to directly use a rigorous technique for deriving the evolution equation for the expectation value of $\vect{\jacobi}$.

It is tempting to conclude that the average angular diameter distance~$\ev{D\e{A}}$ satisfies Eq.~\eqref{eq:evolution_ev_Jacobi} as well, but such an assertion would be wrong, because $D\e{A}=\sqrt{\det\vect{\jacobi}}$ is a nonlinear function of the components of the Jacobi matrix.

\subsubsection{Order-two moments}

We apply the same method to get evolution equations for the order-two moments of $\vect{\jacobi}$. This leads to the following closed system of equations
\begin{align}
\label{eq:syst_second-order_moments_1}
\ddf{}{v}\ev{ \jacobi_{AB} \jacobi_{CD} } &= \ev[1]{\dot{\jacobi}_{AB}\jacobi_{CD}} + \ev[1]{\jacobi_{AB}\dot{\jacobi}_{CD}} \\
\label{eq:syst_second-order_moments_2}
\ddf{}{v}\ev[1]{ \dot{\jacobi}_{AB} \jacobi_{CD} } &= \ev[1]{\dot{\jacobi}_{AB}\dot{\jacobi}_{CD}} 
																						+ \ev{\Ricfoc}\ev{\jacobi_{AB}\jacobi_{CD}}\\
\ddf{}{v}\ev[1]{ \dot{\jacobi}_{AB} \dot{\jacobi}_{CD} } &= \ev{\Ricfoc}\pa{ \ev[1]{\dot{\jacobi}_{AB}\jacobi_{CD}} 
																														+ \ev[1]{\jacobi_{AB}\dot{\jacobi}_{CD}} }
																								+ C_\Ricfoc \ev{\jacobi_{AB}\jacobi_{CD}} \nonumber\\
\label{eq:syst_second-order_moments_3}
								&\qquad + C_\Weylfoc \pa{ \delta_{AC}\delta_{EF} - \eps_{AC} \eps_{EF} }\ev{ \jacobi_{EB}\jacobi_{FD} }
\end{align}
which consists of $10+16+10= 36$ independent equations for the quantities~$\ev{\jacobi_{AB}\jacobi_{CD}}$, $\ev[1]{\dot{\jacobi}_{AB} \jacobi_{CD}}$ and $\ev[1]{\dot{\jacobi}_{AB}\dot{\jacobi}_{CD}}$. By combining the second derivative of Eq.~\eqref{eq:syst_second-order_moments_1} with the derivative of Eq.~\eqref{eq:syst_second-order_moments_2} and Eq.~\eqref{eq:syst_second-order_moments_3}, we can eliminate the moments $\ev[1]{\dot{\jacobi}_{AB} \jacobi_{CD}}$ and $\ev[1]{\dot{\jacobi}_{AB}\dot{\jacobi}_{CD}}$, in order to end up with a closed system for $\ev{\jacobi_{AB}\jacobi_{CD}}$,
\begin{multline}
\ddf[3]{}{v}\ev{\jacobi_{AB}\jacobi_{CD}} = 4\ev{\Ricfoc} \ddf{}{v}\ev{\jacobi_{AB}\jacobi_{CD}} 
																		+ 2 \pa{ \ddf{\ev{\Ricfoc}}{v} + C_\Ricfoc }\ev{\jacobi_{AB}\jacobi_{CD}} \\
																		+ 2 C_\Weylfoc \pa{ \delta_{AC}\delta_{EF} - \eps_{AC} \eps_{EF} }\ev{ \jacobi_{EB}\jacobi_{FD} },
\label{eq:second_order_moments}
\end{multline}
which consists of 10 independent third-order differential equations. We shall not try to solve this system, but rather extract from it information on the angular distance.

\subsubsection{Application to the squared angular distance}\label{sec:squared_angular_distance}

The square of the angular distance is the determinant of $\vect{\jacobi}$, hence quadratic in its components. Its expectation value,
\begin{equation}
\ev{D\e{A}^2} \define \ev{\det\vect{\jacobi}} = \ev{ \jacobi_{11}\jacobi_{22} } - \ev{\jacobi_{12}\jacobi_{21}},
\end{equation}
is therefore ruled by Eq.~\eqref{eq:second_order_moments}. Applying it for $ABCD=1122$ and $ABCD=1221$, we have
\begin{align}
\ddf[3]{}{v}\ev{\jacobi_{11}\jacobi_{22}} &= 4\ev{\Ricfoc} \ddf{}{v}\ev{\jacobi_{11}\jacobi_{22}} 
																		+ 2 \pa{ \ddf{\ev{\Ricfoc}}{v} + C_\Ricfoc }\ev{\jacobi_{11}\jacobi_{22}}
																		- 2 C_\Weylfoc \ev{ D\e{A}^2 }, \\
\ddf[3]{}{v}\ev{\jacobi_{12}\jacobi_{21}} &= 4\ev{\Ricfoc} \ddf{}{v}\ev{\jacobi_{12}\jacobi_{21}} 
																		+ 2 \pa{ \ddf{\ev{\Ricfoc}}{v} + C_\Ricfoc }\ev{\jacobi_{12}\jacobi_{21}}
																		+ 2 C_\Weylfoc \ev{ D\e{A}^2 },
\end{align}
which, by subtraction, yields the following equation for $\ev{D\e{A}^2}$ \emph{only},
\begin{empheq}[box=\fbox]{equation}
\ddf[3]{\ev{D\e{A}^2}}{v} = 4\ev{\Ricfoc} \ddf{\ev{D\e{A}^2}}{v}	+ 2 \pa{ \ddf{\ev{\Ricfoc}}{v} + C_\Ricfoc - 2 C_\Weylfoc}\ev{D\e{A}^2}.
\label{eq:evolution_squared_DA}
\end{empheq}
To our knowledge, it is the first time that such a general exact equation for the evolution of the dispersion of the angular distance in an inhomogeneous universe is derived.

Solving this differential equation requires initial conditions for $\ev{D\e{A}^2}$ and its first and second derivatives. They are easily obtained from the Taylor expansion~\eqref{eq:expansion_DA_near_obs} of $D\e{A}$ for $v\rightarrow 0$,
\begin{equation}
\ev{D\e{A}^2}(0)=0,
\qquad
\ddf{\ev{D\e{A}^2}}{v}(0)=0,
\qquad
\ddf[2]{\ev{D\e{A}^2}}{v}(0)=2.
\end{equation}

Equation~\eqref{eq:evolution_squared_DA} can also be elegantly rewritten in terms of a variable~$x$ defined by
\begin{equation}\label{eq:dx_dv}
\dd x \define \frac{\dd v}{D^2_0(v)},
\end{equation}
where $D_0(v)$ is the background angular distance, i.e. satisfying $\ddot{D}_0=\ev{\Ricfoc}D_0$. It is indeed straightforward to show that the differential operator involved in Eq.~\eqref{eq:evolution_squared_DA} reads
\begin{equation}\label{eq:dx3}
\ddf[3]{}{v} - 4\ev{\Ricfoc} \ddf{}{v} + 2 \ddf{\ev{\Ricfoc}}{v} = D_0^{-4} \ddf[3]{}{x} D_0^{-2}
\end{equation}
so that
\begin{equation}
\ddf[3]{}{x}\pa[4]{\frac{\ev{D\e{A}^2}}{D_0^2}} = 2D_0^4\pa{ C_\Ricfoc - 2 C_\Weylfoc } \ev{ D\e{A}^2 }.
\label{eq:formal_evolution_squared_DA}
\end{equation}
Though formally simpler, this alternative form of Eq.~\eqref{eq:evolution_squared_DA} cannot be used for numerical integration, because $x$ is singular at the observation event---$D_0(v_\obs)=0$---usually chosen as initial condition.

\subsubsection{Expectation value of a general function}

More generally, by multiplying the FPK equation with an arbitrary function~$F(\vect{\jacobi},\vect{\dot{\jacobi}})$ and integrating the right-hand side by parts, we obtain
\begin{multline}
\ddf{\ev{F}}{v} = \ev{ \dot{\jacobi}_{AB} \pd{F}{\jacobi_{AB}} }
							+ \ev{\Ricfoc} \ev{ \jacobi_{AB} \frac{\partial F}{\partial \dot{\jacobi}_{AB}} } \\
							+ \frac{1}{2} \ev{ \pac[2]{ C_\Ricfoc \, \delta_{AE} \delta_{CF}
																+ C_\Weylfoc \pa{\delta_{AC} \delta_{EF} - \eps_{AC}\eps_{EF}}
																	} 
															\jacobi_{EB} \jacobi_{FD} \,
															\frac{\partial^2 F}{\partial\dot{\jacobi}_{AB}\partial\dot{\jacobi}_{CD}}
														}.
\label{eq:ev_arbitrary_function_Jacobi}
\end{multline}
If $F$ is an order-$n$ monomial of the form $F=\jacobi^p \dot{\jacobi}^q$, with $p+q=n$, and where $\jacobi^p$ stands for any product of $p$ components of the Jacobi matrix, then the left-hand side of Eq.~\eqref{eq:ev_arbitrary_function_Jacobi} is $\dd\ev[1]{\jacobi^p \dot{\jacobi}^q}/\dd v$, while the three terms on the right-hand side are respectively of the form $\ev[1]{\jacobi^{p-1}\dot{\jacobi}^{q+1}}$, $\ev[1]{\jacobi^{p+1}\dot{\jacobi}^{q-1}}$, and $\ev[1]{\jacobi^{p+2}\dot{\jacobi}^{q-2}}$, so they are all order-$n$ moments. This confirms what we claimed in the introduction of this section, namely that order-$n$ moments form a closed system of differential equations.

\subsection{Moments of the optical-scalar distribution}\label{sec:moments_scalars}

Contrary to the Jacobi matrix, the optical scalars satisfy a nonlinear Langevin equation. An important consequence on the associated FPK equation~\eqref{eq:FPK_scalar} is that it generates an infinite hierarchy of evolution equations for the moments of the distribution $p(v;\vect{S})$. For instance, if one is interested in computing the average angular distance $\ev{D\e{A}}$, then Eq.~\eqref{eq:FPK_scalar} generates (using the same technique as in \S~\ref{sec:moments_Jacobi})
\begin{align}
\ddf{}{v}\ev{D\e{A}} &= \ev[2]{\nul{\theta} D\e{A}} 
\label{eq:evolution_mean_DA}\\
\ddf{}{v}\ev[2]{\nul{\theta} D\e{A}} &= -\ev[2]{\abs{\nul{\sigma}}^2 D\e{A}} + \ev{\Ricfoc}\ev{D\e{A}}
\label{eq:evolution_mean_thetaDA} \\
\ddf{}{v}\ev[2]{\abs{\nul{\sigma}}^2 D\e{A}} &= -3\ev[2]{ \abs{\nul{\sigma}}^2 \nul{\theta} D\e{A}} + 2 C_\Weylfoc \ev{D\e{A}},
\label{eq:evolution_mean_sigma2DA}\\
\cdots\nonumber
\end{align}
where the evolution of an order-$n$ moment systematically involves order-$(n+1)$ moments. Clearly, such a system cannot be solved analytically, and requires a perturbative approach to be dealt with. A first possibility consists postulating a closure relation for the hierarchy at a given order, but such a method does not seem particularly adapted to the present situation, because the physical meaning of the underlying approximation is unclear, and therefore poorly controlled.

We choose instead to perform a perturbative expansion with respect to the shear rate~$\sigma$, that we assume to be a small quantity. In the following, we focus on the average angular diameter distance~$\ev{D\e{A}}$, and determine its evolution at first and second order in $\abs{\sigma}^2$.

\subsubsection{First-order perturbative expansion}\label{sec:optical_scalars_order1}

We decompose the angular distance and the expansion scalar as
\begin{align}
D\e{A} &= D_0 + D_1, \\
\theta &= \theta_0 + \theta_1,
\end{align}
where $D_0$, already introduced in~\S~\ref{sec:squared_angular_distance}, is the solution of $\ddot{D}_0=\ev{\Ricfoc}D_0$, and $\theta_0\define \dot{D}_0/D_0$ is the corresponding expansion rate; both are deterministic quantities. We assume that the stochastic quantities~$D_1,\theta_1$ are small, in the sense that their probability distributions are concentrated on values much smaller than $D_0,\theta_0$ respectively.

We then expand Eq.~\eqref{eq:evolution_mean_DA} and the following two equations, generated by FPK,
\begin{align}
\ddf{}{v} \ev{\theta} &= -\ev{\theta^2} - \ev[1]{\abs{\sigma}^2} + \ev{\Ricfoc}, 
\label{eq:evolution_mean_theta}\\
\ddf{}{v} \ev[2]{\abs{\sigma}^2} &= -4\ev[1]{\theta \abs{\sigma}^2} + 2C_\Weylfoc,
\label{eq:evolution_mean_sigma2}
\end{align}
at first order in $D_1,\theta_1,\abs{\sigma}^2$, which gives
\begin{align}
\ddf{\ev{D_1}}{v} &= \theta_0 \ev{D_1} + \ev{\theta_1} D_0 \\
\ddf{\ev{\theta_1}}{v} &= -2\theta_0\ev{\theta_1} - \ev[1]{\abs{\sigma}^2} \\
\ddf{\ev[1]{\abs{\sigma}^2}}{v} &= -4\theta_0 \ev[1]{\abs{\sigma}^2} + 2C_\Weylfoc,
\label{eq:evolution_mean_shear_order1}
\end{align}
whence
\begin{empheq}[box=\fbox]{equation}\label{eq:correction_DA_order1}
\delta_{D\e{A}}^{(1)} \define \frac{\ev{D_1}}{D_0}
=
- 2 \int_0^v \frac{\dd v_1}{D^2_0(v_1)}
		\int_0^{v_1} \frac{\dd v_2}{D^2_0(v_2)} 
			\int_0^{v_2}\dd v_3 \; D_0^4(v_3) C_\Weylfoc(v_3)<0,
\end{empheq}
which represents the relative correction between $\ev{D\e{A}}$ and $D_0$ at first order in Weyl lensing. Note that, again, the above result naturally exhibits the integration measure $D_0^{-2}\dd v=\dd x$, it can therefore be rewritten as
\begin{equation}
\ddf[3]{\delta_{D\e{A}}^{(1)}}{x} = - 2 D_0^6 C_\Weylfoc .
\label{eq:formal_evolution_DA}
\end{equation}

\subsubsection{The shear rate at first order}

From Eq.~\eqref{eq:evolution_mean_shear_order1} we have deduced the following expression for the variance of the shear rate,
\begin{equation}\label{eq:variance_shear_order1}
\ev[1]{\abs{\sigma}^2} = 2 \int_0^v \dd w \pac{\frac{D_0(w)}{D_0(v)}}^4 C_\Weylfoc(w),
\end{equation}
at first order. Let us simply mention that, to this order of approximation, we can easily obtain the full PDF of~$\sigma$. Linearizing the second scalar Sachs equation~\eqref{eq:evolution_shear}, we indeed get
\begin{equation}
\dot{\nul{\sigma}} = - 2\nul{\theta}_0 \nul{\sigma} + \Weylfoc + \mathcal{O}(\sigma^2).
\label{eq:linearized_Sachs_shear}
\end{equation}
which is identical to the historical Langevin equation for diffusion.
The associated FPK equation for the PDF~$p_\sigma(v;\sigma)$ is easily shown to be
\begin{equation}
\pd{p_\sigma}{v} 
= 2\nul{\theta}_0 \pa{ \pd{\nul{\sigma}_1p_\sigma}{\nul{\sigma}_1} + \pd{\nul{\sigma}_2p_\sigma}{\nul{\sigma}_2}}
	+ \frac{C_\Weylfoc}{2} \pa{ \pd[2]{p_\sigma}{\nul{\sigma}_1} + \pd[2]{p_\sigma}{\nul{\sigma}_2} }.
\end{equation}
It can be solved by (i) using a polar description for $\nul{\sigma}=\nul{\sigma}_1+\i\nul{\sigma}_2=|\nul{\sigma}|\ex{\i\phi}$, then (ii) using the statistical isotropy assumption that implies $p_\sigma(v;\nul{\sigma}_1;\nul{\sigma}_2)=f(v,|\nul{\sigma}|)$, and (iii) performing simple changes of variable to recover a standard diffusion equation. The result is a Gaussian distribution, describing a 2-dimensional random walk with nonconstant diffusion coefficient,
\begin{equation}\label{eq:PDF_shear_order1}
p_\sigma(v;\sigma) =  \frac{1}{\pi\ev[1]{\abs{\sigma}^2}(v)} \, 
												\exp\pa{ -\frac{\abs{\nul{\sigma}}^2}{\ev[1]{\abs{\sigma}^2}(v)} }
\end{equation}
where $\ev[1]{\abs{\sigma}^2}$ is given by Eq.~\eqref{eq:variance_shear_order1}.

\subsubsection{Second-order perturbative expansion}

In \S~\ref{sec:squared_angular_distance} we derived an evolution equation for~$\ev[1]{D\e{A}^2}$, while \S~\ref{sec:optical_scalars_order1} provided an expression for $\ev{D\e{A}}$. Subtracting the results should therefore lead to the variance of the angular diameter distance. However, the first-order expansion performed in the previous paragraphs is not sufficient for that purpose. This can be understood the following way: if $D\e{A}=D_0+\delta D$, then
\begin{equation}
\varDA\define\ev{D\e{A}^2} - \ev{D\e{A}}^{2} = \ev{\delta D^2} - \ev{\delta D}^2
\end{equation}
involves second-order quantities, neglected in \S~\ref{sec:optical_scalars_order1}. In this paragraph, we therefore expand the equations governing the evolution of $\ev{D\e{A}}$ up to second order in $\ev[1]{\abs{\sigma}^2}$, i.e. formally up to second order in $C_\Weylfoc$.

We start back from Eqs.~\eqref{eq:evolution_mean_DA}, \eqref{eq:evolution_mean_thetaDA}, \eqref{eq:evolution_mean_sigma2DA}, which can be gathered as
\begin{equation}
\ddf[3]{}{x} \pa{ \frac{\ev{D\e{A}}}{D_0} } = -2C_\Weylfoc D_0^5 \ev{D\e{A}} 
																			+ 3D_0^5 \ev[2]{ \abs{\nul{\sigma}}^2 D\e{A} (\theta-\theta_0) }.
\end{equation}
The difficulty now consists in evaluating the last term. First note that, since it is already a second-order quantity,
\begin{equation}
\ev[2]{ \abs{ \nul{\sigma}}^2 D\e{A} (\theta-\theta_0) }
=
D_0 \ev[2]{ \abs{ \nul{\sigma}}^2 \nul{\theta}_1 } + \mathcal{O}(C_\Weylfoc^3).
\end{equation}
Let us then write
\begin{equation}
\ev[2]{ \abs{ \nul{\sigma}}^2 \nul{\theta}_1 }
=
\ev[2]{ \abs{ \nul{\sigma}}^2 } \ev[2]{ \nul{\theta}_1 } 
+ \underbrace{ \ev[2]{ \abs{\nul{\sigma}}^2 \nul{\theta} }  - \ev[2]{ \abs{\nul{\sigma}}^2 } \ev[2]{ \nul{\theta} } }_{\equiv \Gamma_{\theta\sigma}}.
\end{equation}
The first term on the right-hand side can be expressed using the first-order results of \S~\ref{sec:optical_scalars_order1}, which, using the $x$ variable, take the simple form
\begin{align}
\ev[2]{ \abs{ \nul{\sigma}}^2 } &= - D_0^{-4} \ddf[2]{\delta_{D\e{A}}^{(1)}}{x} + \mathcal{O}(C_\Weylfoc^2), 
\label{eq:first-order_correction_distance_shear}\\
\ev[2]{ \nul{\theta}_1 } &= D_0^{-2}\ddf{\delta_{D\e{A}}^{(1)}}{x} + \mathcal{O}(C_\Weylfoc^2).
\end{align}

Evaluating the cross-correlation term~$\Gamma_{\theta\sigma}$ can be achieved by using again the hierarchy of moments generated by the FPK equation. Combining Eqs.~\eqref{eq:evolution_mean_theta}, \eqref{eq:evolution_mean_sigma2} with
\begin{equation}
\ddf{}{v} \ev[2]{ \nul{\theta} \abs{\nul{\sigma}}^2 } = -5\ev[2]{ \nul{\theta}^2 \abs{\nul{\sigma}}^2 } 
																									- \ev[2]{ \abs{\nul{\sigma}}^4 }
																									+ \ev{\Ricfoc} \ev[2]{ \abs{\nul{\sigma}}^2 }
																									+ 2 C_\Weylfoc \ev[2]{ \nul{\theta} },
\end{equation}
we get
\begin{equation}
\dot{\Gamma}_{\theta\sigma} + 6\theta_0 \Gamma_{\theta\sigma} 
= \ev[2]{ \abs{\nul{\sigma}}^2 }^2 - \ev[2]{ \abs{\nul{\sigma}}^4 } + \mathcal{O}(C_\Weylfoc^3),
\label{eq:evolution_cross-correlation_expansion_shear}
\end{equation}
where we have expanded the higher-order correlator~$\ev[1]{ \nul{\theta}^2 \abs{\nul{\sigma}}^2 }$ as $\nul{\theta}_0^2\ev[1]{ \abs{\nul{\sigma}}^2 } + 2 \nul{\theta}_0 \ev[1]{ \nul{\theta}_1 \abs{\nul{\sigma}}^2 } + \mathcal{O}(C_\Weylfoc)$. Now, by comparing the evolution equations for $\ev[1]{ \abs{\nul{\sigma}}^2 }^2$ and $\ev[1]{ \abs{\nul{\sigma}}^4 }$, which are
\begin{align}
\ddf{}{v}\ev[2]{ \abs{\nul{\sigma}}^2 }^2 &= - 8\ev[2]{ \nul{\theta} \abs{\nul{\sigma}}^2 }\ev[2]{\abs{\nul{\sigma}}^2}
																					+ 4 C_\Weylfoc \ev[2]{\abs{\nul{\sigma}}^2}, \\
\ddf{}{v}\ev[2]{ \abs{\nul{\sigma}}^4 } &= - 8\ev[2]{ \nul{\theta} \abs{\nul{\sigma}}^4 }
																					+ 8 C_\Weylfoc \ev[2]{\abs{\nul{\sigma}}^2},
\end{align}
we conclude that $\ev[1]{\abs{\nul{\sigma}}^4}= 2\ev[1]{ \abs{\nul{\sigma}}^2 }^2$ at leading order. Note that this result coincides with the predictions of the Gaussian distribution~\eqref{eq:PDF_shear_order1} obtained for $\sigma$ in the previous paragraph. Hence Eq.~\eqref{eq:evolution_cross-correlation_expansion_shear} is solved as
\begin{align}
\Gamma_{\theta\sigma} &= - D_0^{-6} \int_0^v \dd w \; D_0^6 \ev[2]{ \abs{\nul{\sigma}}^2 }^2 + \mathcal{O}(C_\Weylfoc^3)
\\
										&= - D_0^{-6} \int_\obs^x \dd x' \; \pa{ \ddf[2]{\delta_{D\e{A}}^{(1)}}{x} }^2 + \mathcal{O}(C_\Weylfoc^3),
\end{align}
where we used Eq.~\eqref{eq:first-order_correction_distance_shear}. The lower bound ``$\obs$'' of the latter integral is formal, because variable $x$ is singular for $v=0$. This was the last missing piece to the differential equation governing the evolution of $\ev{D\e{A}}$ at second order in $C_\Weylfoc$,
\begin{equation}\label{eq:evolution_mean_DA_order2}
\ddf[3]{}{x} \pa{ \frac{\ev{D\e{A}}}{D_0} } + 2C_\Weylfoc D_0^6 \, \frac{\ev{D\e{A}} }{D_0}
=	-3 \ddf{\delta_{D\e{A}}^{(1)}}{x} \ddf[2]{\delta_{D\e{A}}^{(1)}}{x} 
	- 3 \int_\obs^x \dd x' \; \pa{ \ddf[2]{\delta_{D\e{A}}^{(1)}}{x} }^2 + \mathcal{O}(C_\Weylfoc^3).
\end{equation}
In terms of an expansion of the form $D\e{A}=D_0+D_1+D_2$, and defining the second-order mean correction~$\delta_{D\e{A}}^{(2)}\define \ev{D_2}/D_0$ to the angular distance, the above result reads
\begin{equation}
\ddf[3]{\delta_{D\e{A}}^{(2)}}{x}
=
-3 \ddf{\delta_{D\e{A}}^{(1)}}{x} \ddf[2]{\delta_{D\e{A}}^{(1)}}{x}
 - 3 \int_\obs^x \dd x' \; \pa{ \ddf[2]{\delta_{D\e{A}}^{(1)}}{x} }^2
<0.
\end{equation}

\subsection{Variance of the angular distance}\label{sec:variance_angular_distance}

We now have enough material to propose an approximate evolution equation for the variance of the angular distance. On the one hand, we have obtained in \S~\ref{sec:moments_Jacobi} the following exact equation for $\ev{D\e{A}^2}$,
\begin{equation}
\ddf[3]{}{x}\pa[4]{ \frac{\ev{D\e{A}^2}}{D_0^2} } + 2D_0^6\pa{ 2 C_\Weylfoc - C_\Ricfoc } \frac{\ev{ D\e{A}^2 }}{D_0^2} = 0.
\label{eq:evolution_DA2_appendix}
\end{equation}
On the other hand, the second-order Eq.~\eqref{eq:evolution_mean_DA_order2} is easily turned into an equation for $\ev{D\e{A}}^2$,
\begin{equation}
\ddf[3]{}{x} \pa{ \frac{\ev{D\e{A}}}{D_0} }^2 + 4 C_\Weylfoc D_0^6 \pa{ \frac{\ev{D\e{A}}}{D_0} }^2
=
-6 \int_\obs^x \dd x' \; \pa{ \ddf[2]{\delta_{D\e{A}}^{(1)}}{x} }^2 + \mathcal{O}(C_\Weylfoc^3).
\end{equation}
By subtraction, we finally obtain
\begin{empheq}[box=\fbox]{equation}
\ddf[3]{}{x}\pac{ \frac{\varDA}{D_0^2} } + 2 D_0^6(2C_\Weylfoc - C_\Ricfoc) \frac{\varDA}{D_0^2}
=
2 C_\Ricfoc D_0^6 
+ 6 \int_\obs^x \dd x' \pac{\ddf[2]{\delta^{(1)}_{D\e{A}}}{x}}^2
+ \mathcal{O}(C_\Weylfoc^3),
\label{eq:evolution_variance_DA}
\end{empheq}
where we recall that $\dd x=D_0^{-2} \dd v$, and that the third derivative $\dd^3/\dd x^3$ is given by Eq.~\eqref{eq:dx3}. We see that both Ricci lensing and Weyl lensing drive the variance of $D\e{A}$. This can be easily understood from the focusing theorem~\eqref{eq:focusing_theorem}, where $\Ricfoc$ is the main driving term, which explains why $C_\Ricfoc$ appears directly on the right-hand side of \eqref{eq:evolution_variance_DA}; $\Weylfoc$, on the other hand, affects $D\e{A}$ only indirectly, via $\abs{\sigma}^2$. It is the reason why $\dd^2\delta_{D\e{A}}^{(1)}/\dd x^2\propto\ev[1]{\abs{\sigma}^2}$ is also present on the right-hand side of Eq.~\eqref{eq:evolution_variance_DA}.

It is remarkable that this result on the variance of $D\e{A}$ required the use of both the Jacobi matrix and the optical scalars. Although they are completely equivalent formulations, it would have been much more painful to derive Eq.~\eqref{eq:evolution_variance_DA} by using exclusively one of them.

\section{Application to a Swiss-cheese model}\label{sec:application_SC}

The stochastic lensing formalism developed throughout Secs.~\ref{sec:Langevin}, \ref{sec:FPK}, and \ref{sec:general_results} depends on three free functions: the average Ricci focusing~$\ev{\Ricfoc}(v)$, and the two covariances amplitudes~$C_\Ricfoc(v)$, $C_\Weylfoc(v)$ which need to be specified, or deduced from a spacetime model, in order to draw any physical conclusion. In this section, we propose an application of this formalism to Swiss-cheese (SC) cosmological models. Our goal is twofold: on the one hand, it provides an explicit example about how stochastic lensing can be applied, and of the involved calculations; on the other hand, it allows us to test its validity, by comparing its analytical predictions with the numerical results of a ray-tracing code for SC models, which was developed by one of the authors and used in Refs.~\cite{Fleury:2013sna,Fleury:2014gha}. As a byproduct, we also obtain an improvement of the Kantowski-Dyer-Roeder approximation, which allows for shear.

\subsection{The Einstein-Straus Swiss-cheese model}\label{geometry_SC}

We consider here an Einstein-Straus~\cite{SW1,SW2,SW3,SW4,2013GReGr..45.2143M} SC model, where individual masses, whose vicinity is characterised by the Schwarzschild solution (or the Kottler solution, for a nonvanishing cosmological constant), are embedded in an expanding homogeneous and isotropic Universe, forming spherical holes within the Friedmannian cheese. This model aims at describing static, gravitationally bound objects, such as stars, galaxies, or clusters of galaxies, and is therefore more adapted to the problematic of small-scale inhomogeneities tackled here than LTB~\cite{LTB1,inhomSol} or Szekeres~\cite{SZE,inhomSol} Swiss-cheese models.

\subsubsection{Spacetime geometry}

Let us briefly summarise the main geometrical properties of the Einstein-Straus model---more detailed explanations can be found, e.g., in our previous works~\cite{Fleury:2013sna,Fleury:2014gha}. Consider one hole of the SC, whose centre is taken to be the origin of the coordinate system, without loss of generality. On the one hand, the metric of the exterior region is
\begin{equation}\label{eq:FL_metric}
\dd s^2 = -\dd T^2 + a^2(T)\left[\frac{\dd R^2}{1-K R^2} +R^2\,\dd\Omega^2\right],
\end{equation}
with $\dd\Omega^2\define\dd\theta^2 + \sin^2\theta \dd\ph^2$, $K=\mathrm{cst}$, and where the evolution of the scale factor~$a$ with cosmic time~$T$ is ruled by the Friedmann equations, in particular
\begin{equation}
H^2 \define \pa{\frac{1}{a} \ddf{a}{T} }^2 = \frac{8\pi G \rho_0}{3} \pa{\frac{a_0}{a}}^3 - \frac{K}{a^2} + \frac{\Lambda}{3},
\end{equation}
where $\rho_0$ is today's mean density of matter, modelled by a pressureless fluid. The cosmological parameters quantifying the relative importance of matter, spatial curvature, and cosmological constant in the expansion dynamics are respectively  $\Omega\e{m}\define 8\pi G\rho_0/(3 H^2)$, $\Omega_{K}\define -K/(a H)^2$, and $\Omega_\Lambda\define \Lambda/(3H^2)$. The interior geometry is, on the other hand, given by the Kottler (or Schwarzschild-de Sitter) metric
\begin{equation}\label{eq:Kottler_metric}
\dd s^2 = -A(r) \, \dd t^2 + A^{-1}(r) \, \dd r^2 + r^2\dd\Omega^2
\qquad\text{with} \quad 
A(r) \equiv 1 - \frac{r\e{S}}{r} - \frac{\Lambda\,r^2}{3},
\end{equation}
and where $r\e{S}\equiv 2\,GM$ is the Schwarzschild radius associated with the mass $M$ at the centre of the hole.

The metrics~\eqref{eq:FL_metric} and \eqref{eq:Kottler_metric} are glued together on a spacelike hypersurface corresponding a comoving sphere (the boundary of the hole), hence defined by $R=R\e{h}=\mathrm{cst}$ in terms of exterior coordinates, and $r=r\e{h}(t)$ in terms of interior coordinates. The Darmois-Israel junction conditions~\cite{Darmois,1966NCimB..44....1I,1967NCimB..48..463I} then impose
\begin{align}
r\e{h}(t) &= a(T) R\e{h}, \\
M &= \frac{4\pi}{3} \rho_0 R\e{h}^3. \label{eq:junction_2}
\end{align}
Equation~\eqref{eq:junction_2} must be understood as follows: the mass~$M$ at the centre of the hole is identical to the one that should be contained in the sphere of comoving radius~$R\e{h}$, if the latter were homogeneously filled with the same comoving density~$\rho_0$ as the exterior.

\subsubsection{Optical properties of each region}

Within the cheese, since the FL metric is conformally flat, light rays follow straight lines in terms of a suitable coordinate system. The cyclic frequency of the associated wave, as measured by a comoving observer (with four-velocity $\partial_T$), and normalised by the observed frequency at $O$, reads
\begin{equation}
1+z = \omega = k^T = \ddf{T}{v} = \frac{a_0}{a(T)}
\end{equation}
from which follows the relation between redshift~$z$ and affine parameter~$v$ for a light ray propagating through the cheese only,
\begin{equation}\label{eq:z_v_FL}
\ddf{v}{z} = \frac{1}{H(z)(1+z)^2}.
\end{equation}
Besides, the Ricci and Weyl lensing scalars are shown to be
\begin{align}
\Ricfoc\e{FL} &= -4\pi G \omega^2 \rho(T) \\
\Weylfoc\e{FL} &= 0. 
\end{align}

\begin{figure}[h!]
\centering
\includegraphics[scale=0.8]{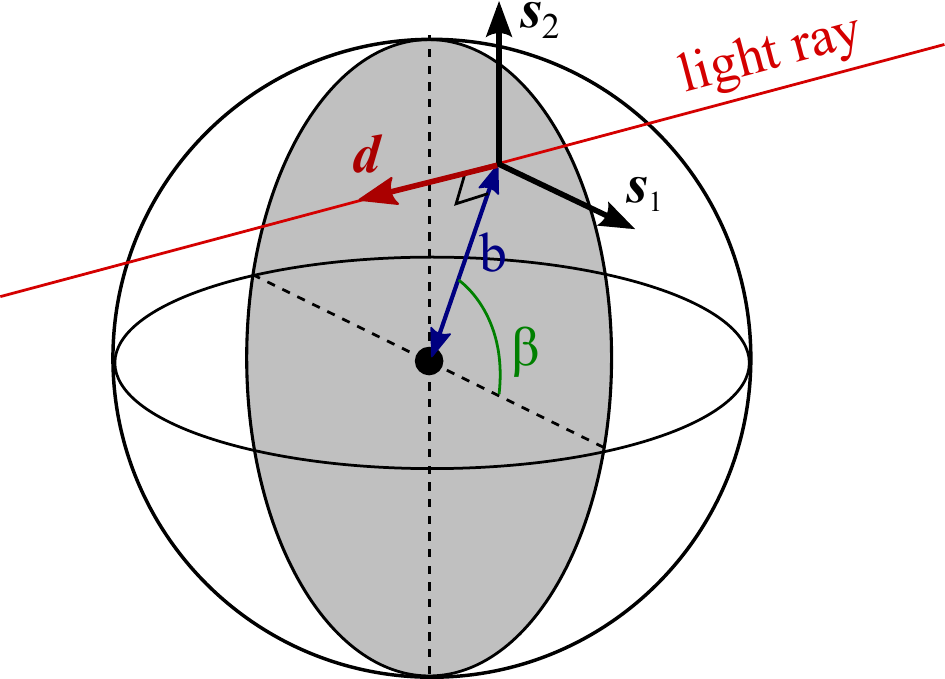}
\caption{Impact parameters in a Kottler hole. The grey disk is the intersection between the hole and the plane orthogonal to the wave-vector~$\vect{k}$ at minimal approach, also spanned by the Sachs basis~$(\vect{s}_1,\vect{s}_2)$ there. The impact parameter~$b\define L/E$ is approximately the minimal approach radial coordinate of the photon, and $\beta$ is the angle between the plane of the trajectory and the plane spanned by $\vect{k}$, $\vect{s}_1$ at minimal approach. It is also the angle corresponding the basis change which diagonalises the optical tidal matrix~$\vect{\optical}\e{K}$ in the hole.}
\label{fig:impact_parameters}
\end{figure}

Inside the hole, a light ray propagating in the $\theta=\pi/2$-plane admits two constants of motion, $E=A(r) k^t$ and $L=r^2 k^\ph$, respectively associated with the stationarity and spherical symmetry of the metric. Their ratio defines the impact parameter $b=L/E$, roughly equal to the closest approach radius $r\e{min} \approx b$ of the photon trajectory if $b\gg r\e{S}$. The Ricci and Weyl lensing scalars read, in this case,
\begin{align}
\Ricfoc\e{K} &= 0 \\
\Weylfoc\e{K} &= \frac{3 GM L^2}{r^5} \, \ex{-2\i\beta}, \label{eq:Weyl_Kottler}
\end{align}
where $\beta$ is the impact angle, corresponding to the angle between the plane of the trajectory and the first vector of the Sachs basis, as represented on Fig.~\ref{fig:impact_parameters}.

\subsection{Effective optical properties}

Because of the intrinsically discrete nature of the SC model, we need to design an effective approach to be able to use the formalism developed in this paper.

\subsubsection{The Kantowski-Dyer-Roeder approximation}\label{sec:KDR}

The first set of effective optical properties for SC models was proposed by Kantowski~\cite{1969ApJ...155...89K} in 1969, assuming that the mass clumps modelled by the central mass of the holes are extended and opaque, i.e., imposing a cutoff for the impact parameter $b>b\e{min}$, which corresponds to the physical radius~$r\e{phys}$ of the clump. This work was generalised in 1974 by Dyer and Roeder~\cite{1974ApJ...189..167D} in order to include the cosmological constant. The resulting behaviour at lowest order, that we shall call the Kantowski-Dyer-Roeder (KDR) approximation, can be summarised as follows:
\begin{description}
\item[KDR1] The relation between affine parameter $v$ and redshift $z$ is not significantly affected by the holes, so that Eq.~\eqref{eq:z_v_FL} can still be applied in a SC model.
\item[KDR2] The effect of the shear, due to Weyl lensing in the holes, on the angular distance is negligible. In other words, $\Weylfoc\e{KDR}=0$.
\item[KDR3] Ricci lensing is the same as in the cheese but reduced by a factor $\bar{\alpha}\in[0,1]$, called smoothness parameter, so that $\Ricfoc\e{KDR}=\bar{\alpha}\Ricfoc\e{FL}=-4\pi G\omega^2\bar{\alpha}\rho(T)$.
\end{description}

A detailed analysis of this approximation was presented recently in Ref.~\cite{Fleury:2014gha}. Hypothesis~\textbf{KDR1} turns out to be valid up to terms on the order of the ratio~$r\e{S}/r\e{h}$ between the Schwarzschild radius of the central mass and the radius of the hole, which is very small in practice. Therefore, \emph{we will adopt \textbf{KDR1} for the remainder of this article}. The relevance of \textbf{KDR3} can be understood as follows: consider an interval~$[v_n,v_{n+1}]$ of the light path, where $v_n$ corresponds to the entrance into the hole number $n$, and $v_{n+1}=v_n+\Delta v_n$ to the entrance into the next one; the effective Ricci focusing over this interval can be defined as
\begin{equation}
\Ricfoc\e{eff} 
\define \frac{1}{\Delta v_n} \int_{v_n}^{v_{n+1}} \Ricfoc \; \dd v
\approx \frac{\Delta v\h{FL}_n}{\Delta v_n} \, \Ricfoc\e{FL},
\end{equation}
where $\Delta v\h{FL}_n$ is the fraction of light path spent into the FL region (between the exit from the hole~$n$ and the entrance into the hole~$n+1$) over which $\Ricfoc\e{FL}$ can be considered constant. This defines a local smoothness parameter~$\alpha_n\define \Delta v\h{FL}_n/\Delta v_n$. Interpolating the sequence $(\alpha_n)$ on the whole light path yields a function~$\alpha(v)$ which, after averaging over many lines of sights, defines $\bar{\alpha}(v)$.

In terms of the stochastic lensing formalism, we can thus identify
\begin{equation}
\ev{\Ricfoc} = \Ricfoc\e{KDR}.
\end{equation}
As a consequence, the angular diameter distance predicted by the KDR approximation corresponds to $D_0$ introduced in \S~\ref{sec:squared_angular_distance}, i.e. satisfying $\ddot{D}_0=\ev{\Ricfoc} D_0$.

\subsubsection{Effective Weyl lensing in a hole}

Numerical ray-tracing simulations in SC models~\cite{Fleury:2014gha} show that, while the KDR approximation satisfactorily reproduces the true $D\e{A}(v)$ relation for most lines of sights, some exhibit significant deviations. Such discrepancies are due to Weyl lensing, neglected in the KDR approach (\textbf{KDR2}), but which we would like to include in the stochastic approach. It will be convenient, for that purpose, to first derive an effective expression for the Weyl lensing scalar~$\Weylfoc$ in a single hole, defined as
\begin{equation}\label{eq:effective_Weyl_hole_def}
\Weylfoc\e{eff} \define \frac{1}{v\e{out}-v\e{in}} \int_{v\e{in}}^{v\e{out}} \Weylfoc\e{K}(v) \; \dd v,
\end{equation}
where $v\e{in}$, $v\e{out}$ respectively denote the affine parameter at entrance and exit.

Like in the KDR approach, we assume from now on that the central mass is an extended opaque object, whose physical radius~$r\e{phys}\gg r\e{S}$ is thus a lower cutoff of impact parameters. As shown in Ref.~\cite{Fleury:2013sna}, the radial coordinate~$r(v)$ of a photon propagating through the hole with an impact parameter~$b$ reads, at lowest order in $r\e{S}/b$,  
\begin{equation}
r(v) \approx \sqrt{b^2+E^2(v-v\e{m})^2},
\end{equation}
where $v\e{m}$ denotes the affine parameter at minimal approach, and $E\approx \omega\e{in}\approx \omega\e{out}$. Moreover, if we neglect the growth of the hole between the photon entrance and exit, then
\begin{equation}
v\e{out}-v\e{m} \approx v\e{m} - v\e{in} \approx E^{-1} \sqrt{r\e{h}^2-b^2}.
\end{equation}
Calculating the integral of Eq.~\eqref{eq:effective_Weyl_hole_def} thus yields
\begin{align}
\Weylfoc\e{eff} &= GM E^2 \pac{ \frac{1}{r\e{h}^3} + \frac{2}{b^2 r\e{h}} } \ex{-2\i\beta} \\
								&= 4\pi G \rho \omega^2 \pac{ \frac{1}{3} + \frac{2}{3} \pa{\frac{r\e{h}}{b}}^2 } \ex{-2\i\beta} .
								\label{eq:effective_Weyl_hole_result}
\end{align}
We see that, for $b\ll r\e{h}$, the ratio between Weyl and Ricci lensing can actually be very large, $|\Weylfoc\e{eff}|/\Ricfoc\e{eff}\propto (r\e{h}/b)^2$. It is the randomization of $\beta$ which, in practice, drastically reduces the net impact of Weyl lensing on the angular distance.

\subsection{Calculation of the covariance amplitudes}\label{sec:covariance_SC}

We now turn to the calculation of the statistical quantities~$C_\Ricfoc$, $C_\Weylfoc$ of the white noises which best reproduce lensing in a Swiss-cheese model.

\subsubsection{Statistical setup}\label{sec:statistical_setup}

The randomness of our SC model is constructed in a way that---as originally formulated by Ref.~\cite{Holz:1997ic}---``each ray creates its own Universe''. One realization of the various stochastic processes at stake thus corresponds to the disposition of successive holes on a photon's trajectory, with random sizes, impact parameters, and separations. Expectation values~$\ev{\ldots}$ will be considered with respect to such realizations. As in Ref.~\cite{Fleury:2014gha}, we make the following assumptions:
\begin{itemize}
\item The properties (mass, size, impact parameters) of two different holes are {independent}, as well as the separation between different successive holes.
\item All the impact positions, within a given hole cross-section, are equiprobable. In other words, the impact angle $\beta$ is uniformly distributed in $[0,2\pi]$, and the PDF of the comoving areal impact parameter~$B$ is
\begin{equation}
p(B)\,\dd B = [R\e{c}\leq B \leq R\e{h}] \, \frac{B\,\dd B}{R\e{h}^2-R\e{c}^2},
\end{equation}
where the squared bracket is $1$ if the assertion inside is true, $0$ if not; $R\e{c}$ denotes the comoving areal radius of the central matter clump, and $R\e{h}$ the comoving areal radius of the hole. We assume that the matter clump is static, i.e., its physical radius $r\e{c}\define a R\e{c}$ is constant, hence $R\e{c}\propto a^{-1}$ is not, contrary to $R\e{h}$.
\item The distributions of both $R\e{h}$ and $r\e{c}$ are governed by the specific matter clumps that one wishes to model. For most of our theoretical results, they do not need to be explicitly specified. For numerical illustrations, we consider galaxylike clumps which all have the same physical density~$\rho\e{c}=3 M/(4\pi r\e{c}^3)=3.47\times 10^{-22}\U{kg/m^3}$---this fixes the relation between $r\e{c}$ and $M$ (hence $R\e{h}$)---, and whose mass function is inspired from Ref.~\cite{2004MNRAS.355..764P},
\begin{equation}
p(M)\dd M \propto M^{-1.16} \exp\pa{-\frac{M}{7.5 \times 10^{11} h^{-2} M_\odot}} \dd M.
\end{equation}
\item The PDF of the comoving separation~$\Delta\chi\e{FL}$ between two successive holes is taken to be uniform, between $0$ and $2\ev{\Delta\chi\e{FL}}$, with
\begin{equation}
\ev{\Delta\chi\e{FL}} = \frac{4}{3}\frac{\bar{\alpha}}{1-\bar{\alpha}} \ev{R\e{h}}.
\end{equation}
This choice ensures that the mean smoothness parameter~$\ev{\alpha}=\ev{\Delta v\e{FL}/\Delta v}$ is indeed~$\bar{\alpha}$.
\end{itemize}

\subsubsection{Ricci-lensing covariance}\label{sec:Ricci_covariance}

In reality, the Ricci and Weyl lensing scalars in a random Swiss-cheese model are not white noises: they have a self-correlation length on the order of the hole sizes. We here aim at determining the properties of the white noises which best reproduce the actual behaviour of $\Ricfoc$ and $\Weylfoc$. In the case of the Ricci covariance amplitude, this can be achieved by integrating Eq.~\eqref{eq:def_covariance_Ricci} with respect to $w$,
\begin{equation}
C_\Ricfoc(v)
= \int \dd w \ev{ \delta\Ricfoc(v) \delta\Ricfoc(w) } 
\approx \int \dd w \ev{ \delta\Ricfoc\e{eff}(v) \delta\Ricfoc\e{eff}(w) },
\end{equation}
with
\begin{equation}
\delta\Ricfoc\e{eff} \define \Ricfoc\e{eff} - \ev{\Ricfoc\e{eff}} 
										= -4\pi G  \rho_0 \, \omega^5 \delta\alpha ,
\end{equation}
and $\delta\alpha(v)\define \alpha(v)-\bar{\alpha}$. As mentioned above, the expectation value~$\ev{\ldots}$ is identified with an average over all possible realizations (r) of the SC, that is over the position, size, and impact parameter of each hole that is crossed by the light beam,
\begin{equation}
\ev{ \delta\Ricfoc\e{eff}(v) \delta\Ricfoc\e{eff}(w) }
=
\lim_{N\to\infty}
\frac{1}{N}\sum_{\rm r=1}^N \delta\Ricfoc\e{eff}\h{(r)}(v) \delta\Ricfoc\e{eff}\h{(r)}(w).
\label{eq:covariance_Ricci_schematic}
\end{equation}

For each realization (r), the complete light path through the SC can be split into elementary intervals $I_n\define[v_n,v_{n+1}]$, of affine parameter length $\Delta v_n$ where, as before, $v_n$ corresponds to the entrance into the $n$th hole. Within each interval, $\delta\Ricfoc\e{eff}=\delta\Ricfoc_n$ is considered constant, and $\delta\Ricfoc_n$ is independent of $\delta\Ricfoc_m$ if $n\not= m$. Hence, if we call $I\e{(r)}(v)$ the elementary interval of (r) such that $v\in I\e{(r)}(v)$, then there are two categories of realizations: those where $w\in I\e{(r)}(v)$ as well; and those where $w\not\in I\e{(r)}(v)$. The net contribution of the second category to the sum of Eq.~\eqref{eq:covariance_Ricci_schematic} vanishes.

In order to calculate this sum, it is convenient to sort the realizations (r) in terms of the properties of $I\e{(r)}(v)$. The affine-parameter length~$\Delta v$ of any elementary interval~$I$ can be decomposed into its FL and hole contributions as
\begin{equation}
\Delta v = \Delta v\e{FL} + \Delta v\e{h} = \frac{1}{\omega^2} \pa{ \Delta\chi\e{h} + 2\sqrt{R\e{h}^2-B^2} },
\end{equation}
where we neglected the global beam deflection in the hole part, and used the FL relation between affine parameter and comoving distance, even in the hole\footnote{This operation is justified by \textbf{KDR1}, which is very accurately satisfied in a SC model }. $\Delta v$ thus depends on the random parameters $\Delta\chi\e{FL}$, $R\e{h}$, and $B$, which we regroup in a triple~$\vect{\Pi}=(\Delta\chi\e{FL},r\e{h},B)$. We now organise the sum of Eq.~\eqref{eq:covariance_Ricci_schematic} in terms of the parameters $\vect{\Pi}$ characterizing the interval containing $v$, which yields 
\begin{equation}\label{eq:covariance_Ricci_formal}
\ev{ \delta\Ricfoc\e{eff}(v) \delta\Ricfoc\e{eff}(w) }
=
\int \dd\vect{\Pi} \; p(\vect{\Pi}| v \in I_{\vect{\Pi}}) \, \prob(w\in I_{\vect{\Pi}}|v \in I_{\vect{\Pi}},\vect{\Pi}) \,
\delta\Ricfoc\e{eff}^2(\vect{\Pi}).
\end{equation}
In the above equation, $p(\vect{\Pi}| v\in I_{\vect{\Pi}})\,\dd\vect{\Pi}$ represents the (conditional) probability that the interval~$I_{\vect{\Pi}}$ containing $v$ has its parameters within~$\dd\vect{\Pi}$ around $\vect{\Pi}$. It can be rewritten thanks to the Bayes formula as
\begin{equation}\label{eq:Bayes}
p(\vect{\Pi}| v\in I_{\vect{\Pi}}) = \frac{\prob(v\in I_{\vect{\Pi}}|\vect{\Pi})}{\prob(v\in I)} \times p(\vect{\Pi}),
\end{equation}
where $p(\vect{\Pi})$ is the unconstrained PDF of $\vect{\Pi}$, i.e. as provided by the assumptions of \S~\ref{sec:statistical_setup}. Simple geometric arguments show that the probability that $v$ belongs to a given interval $I_{\vect{\Pi}}$, with affine-parameter length $\Delta v(\vect{\Pi})$, is
\begin{equation}
\prob(v\in I_{\vect{\Pi}}|\vect{\Pi}) \propto \Delta v,
\end{equation}
so that the normalization factor in the denominator of Eq.~\eqref{eq:Bayes} is simply $\prob(v\in I)\propto \ev{\Delta v}_{\vect{\Pi}}$, where the average is performed with respect to $p(\vect{\Pi})$.

The second term in the integral of Eq.~\eqref{eq:covariance_Ricci_formal} represents the probability that $w$ belongs to the interval~ $I_{\vect{\Pi}}$, given its parameters~$\vect{\Pi}$ and the fact that $v$ already belongs to it. Again, simple geometry yields
\begin{equation}
\prob(w\in I_{\vect{\Pi}}|v \in I_{\vect{\Pi}},\vect{\Pi})
= \pa{ 1 - \frac{\abs{v-w}}{\Delta v} } \Theta( \Delta v - \abs{v-w} ),
\end{equation}
where $\Theta$ denotes the Heaviside function. Gathering all the results, and using the expression of $\delta\Ricfoc\e{eff}$, we obtain
\begin{equation}
\ev{ \delta\Ricfoc\e{eff}(v) \delta\Ricfoc\e{eff}(w) }
=
(4\pi G\rho_0\omega^5)^2
\int \dd\vect{\Pi} \; p(\vect{\Pi}) \,
\frac{\Delta v -\abs{v-w}}{\ev{\Delta v}_{\vect{\Pi}}} \,
\Theta( \Delta v - \abs{v-w} ) \pa{ \frac{\Delta v\e{FL}}{\Delta v} - \bar{\alpha} }^2.
\end{equation}
Performing the integration, plus the one with respect to $w$, finally yields
\begin{empheq}[box=\fbox]{equation}\label{eq:covariance_Ricci_result}
C_\Ricfoc
=
\bar{\alpha}^2(1-\bar{\alpha}) H_0^4 \Omega\e{m0}^2 (1+z)^8
\pa{ \frac{11}{8} \ev{R\e{h}} 
+ \frac{27}{8} \frac{\ev{R\e{h}^2}-\ev{R\e{h}}^2}{\ev{R\e{h}}} }
\end{empheq}
in terms of the usual cosmological quantities. In the above equations, angle brackets denote averaging with respect to the mass function of the matter clumps, which rules the size of the hole they belong to via Eq.~\eqref{eq:junction_2}. Note that we get $C_\Ricfoc=0$ in both limits $\bar{\alpha}=0,1$. This was indeed expected: for $\bar{\alpha}=0$ the Swiss cheese is completely filled by holes, so that $\Ricfoc=0$ everywhere; for $\bar{\alpha}=1$, we recover the strictly homogeneous FL spacetime, in which $\Ricfoc=\ev{\Ricfoc}$ everywhere. In both cases the fluctuation~$\delta\Ricfoc$ vanishes.

\subsubsection{Weyl-lensing covariance}\label{sec:Weyl_covariance}

Just like in the Ricci case, the covariance amplitude~$C_\Weylfoc$ of the white noise which best reproduces Weyl lensing in a SC model is
\begin{equation}
C_\Weylfoc(v) = \frac{1}{2} \int \dd w \ev{ \Weylfoc(v) \Weylfoc^*(w) }
						\approx	\frac{1}{2} \int \dd w \ev{ \abs{\Weylfoc\e{eff}(v) \Weylfoc\e{eff}(w)} \ex{2\i\beta(w)-2\i\beta(v)} },
\end{equation}
where a star denotes the complex conjugate, and the $1/2$ prefactor comes from the fact that in Eq.~\eqref{eq:def_covariance_Weyl} we defined $C_\Weylfoc$ as the covariance amplitude of each independent component $\Weylfoc_A$.

We then proceed as before, decomposing the expectation value~$\ev{ \Weylfoc\e{eff}(v) \Weylfoc\e{eff}^*(w) }$ as a sum over all possible realizations of the SC. Since $\Weylfoc\e{eff}$ is nonzero only in holes, we fully decompose each realization into FL and hole elementary paths (rather that \{FL+hole\} sets as before). In the average, only the realizations such that $v$ and $w$ belong to the same hole~$H$ contribute to the net result. Hence the analogue of Eq.~\eqref{eq:covariance_Ricci_formal} is
\begin{equation}
\ev{ |\Weylfoc\e{eff}(v) \Weylfoc\e{eff}(w)| }
=
(1-\bar{\alpha})
\int \dd\vect{\Pi} \; p(\vect{\Pi}| v \in H_{\vect{\Pi}}) \, \prob(w\in H_{\vect{\Pi}}|v \in H_{\vect{\Pi}},\vect{\Pi})
\, |\Weylfoc\e{eff}(\vect{\Pi})|^2,
\end{equation}
where $\vect{\Pi}$ is now the couple~$(B,R\e{h})$ characterising a hole $H$. The $(1-\bar{\alpha})$ prefactor corresponds to the probability that the elementary interval to which belong $v$ is a hole. The involved probabilities are formally identical to the Ricci case, except that the interval length is now $\Delta v\e{h}$ instead of $\Delta v=\Delta v\e{h}+\Delta v\e{FL}$. The integral to calculate is therefore
\begin{multline}
\ev{ |\Weylfoc\e{eff}(v) \Weylfoc\e{eff}(w)| }
=
(1-\bar{\alpha})(4\pi G\rho_0\omega^5)^2
\int \dd\vect{\Pi} \; p(\vect{\Pi}) \,
\frac{\Delta v\e{h} -\abs{v-w}}{\ev{\Delta v\e{h}}_{\vect{\Pi}}} \, \Theta( \Delta v\e{h} - \abs{v-w} ) \\
\times \pac{ \frac{1}{3} + \frac{2}{3}\pa{\frac{R\e{h}}{B}}^2 }^2.
\end{multline}
The final result, after integration over $\vect{\Pi}$ and $w$, is
\begin{empheq}[box=\fbox]{equation}
C_\Weylfoc = \frac{3}{2} (1-\bar{\alpha}) H_0^2 \Omega\e{m0} (1+z)^6 \,
\frac{\ev{r\e{S}^{4/3} r\e{c}^{-2}}}{\ev{r\e{S}^{1/3}}},
\label{eq:covariance_Weyl_result}
\end{empheq}
Like in Eq.~\eqref{eq:covariance_Ricci_result}, angle brackets denote here averages with respect to the statistical properties of the matter clumps.

\begin{figure}[h!]
\centering
\includegraphics[width=0.5\linewidth]{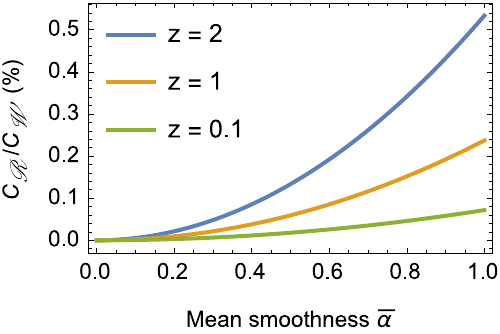}
\caption{Ratio~$C_\Ricfoc/C_\Weylfoc$ between the covariance amplitudes of the Ricci lensing and Weyl lensing in a Swiss-cheese model, as a function of the mean smoothness parameter~$\bar{\alpha}$, for three different values of the redshift~$z=2$ (blue), $z=1$ (orange), and $z=0.1$ (green).}
\label{fig:comparison_CR-CW}
\end{figure}

A comparison of the covariance amplitudes $C_\Ricfoc$ and $C_\Weylfoc$, calculated with the setup and numerical values listed in \S~\ref{sec:statistical_setup}, is depicted in Fig.~\ref{fig:comparison_CR-CW}. It is clear here that Weyl covariance dominates over Ricci covariance. This result is characteristic of the Einstein-Straus SC model, where the local matter density experienced by light oscillates between $\rho$ (cheese) and $0$ (holes); this \emph{highly underestimates} the fluctuations of Ricci focusing compared to reality.

\subsection{Results and comparison with ray tracing}\label{sec:results_SC}

We now apply the general results derived in Sec.~\ref{sec:general_results} with the expressions~\eqref{eq:covariance_Ricci_result} and \eqref{eq:covariance_Weyl_result} for $C_\Ricfoc$ and $C_\Weylfoc$. After having discussed our expression of the average shear rate with respect to earlier works, we compare the predictions of our formalism for $\ev{D\e{A}}$ and $\varDA$ with the output of numerical ray-tracing simulations in a SC model.

\subsubsection{Shear rate and astrophysical parameter}

\begin{figure}[h!]
\centering
\includegraphics[width=0.5\linewidth]{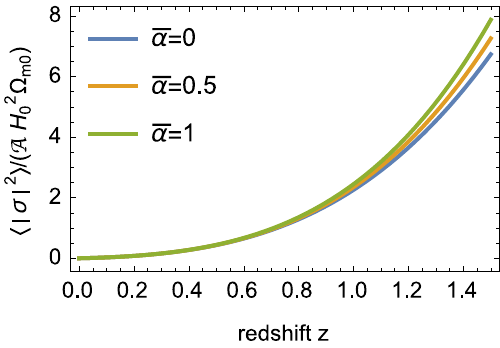}
\caption{Evolution of the integral of Eq.~\eqref{eq:variance_shear_SC}, as a function of the redshift, for three different smoothness parameters $\bar{\alpha}=0,0.5,1$.}
\label{fig:shear_redshift}
\end{figure}

Introducing the expression~\eqref{eq:covariance_Weyl_result} of $C_\Weylfoc$ in Eq.~\eqref{eq:variance_shear_order1} yields the following formula for the average shear rate
\begin{equation}
\ev[2]{ \abs{\sigma}^2 }
=
\mathcal{A} \, H_0^3 \Omega\e{m0}
\int_0^v \dd w \pac{\frac{D_0(w)}{D_0(v)}}^4 (1+z)^6,
\label{eq:variance_shear_SC}
\end{equation}
where we introduced a dimensionless \emph{astrophysical parameter}
\begin{equation}\label{eq:astrophysical_parameter}
\mathcal{A}
\define
\frac{3}{H_0} (1-\bar{\alpha}) \, \frac{\ev{r\e{S}^{4/3} r\e{c}^{-2}}}{\ev{r\e{S}^{1/3}}},
\end{equation}
which encodes the statistical assumptions about the mass and compacity of the matter clumps. It also contains the main dependence with respect to the smoothness parameter~$\bar{\alpha}$, since the integral of Eq.~\eqref{eq:variance_shear_SC} is almost independent from it, as shown in Fig.~\ref{fig:shear_redshift}.
In terms of orders of magnitude, for $\bar{\alpha}=0$, $\mathcal{A}_0\sim g/H_0$, where $g\define GM/r\e{c}^2$ is the surface gravity of the central matter clumps. If they represent galaxies, then $\mathcal{A}_0$ is typically of order unity, but it is potentially much larger for more compact objects (see table~\ref{tab:astro_parameter}).

\begin{table}[h!]
\centering
\begin{tabular}{ccccc}
\hline 
\hline
Nature of the clumps & $M$ & $r\e{S}$ & $r\e{c}$ & $\mathcal{A}_0$ \\ 
\hline
galaxy clusters & $10^{15} M_\odot$ & $100\U{pc}$ & $10\U{Mpc}$ & $10^{-3}$  \\ 
galaxies & $10^{11} M_\odot$ & $10^{-2}\U{pc}$ &$10\U{kpc}$ & $1$ \\ 
stars & $M_\odot$ & km & $10^6\U{km}$ & $10^{10}$ \\ 
\hline 
\hline
\end{tabular} 
\caption{Typical orders of magnitude for the mass~$M$, Schwazschild radius~$r\e{S}$, and physical size~$r\e{c}$ of three possible types of matter clumps modelled in a SC model, with the associated astrophysical parameter~$\mathcal{A}_0\sim r\e{S}/(H_0 r\e{c}^2)$ for $\bar{\alpha}=0$.}
\label{tab:astro_parameter}
\end{table}

Equation~\eqref{eq:variance_shear_SC} is very similar to the ones obtained, e.g., by Gunn~\cite{1967ApJ...150..737G} or Kantowski~\cite{1969ApJ...155...89K} by different methods. Both get the same integral term, but their estimations of the astrophysical parameter differ with ours. In particular, Kantowski obtains\footnote{Dyer and Roeder also obtained the same result, given in Eq.~(25) of Ref.~\citep{1974ApJ...189..167D} with no derivation, but referring to Dyer's PhD thesis~\cite{1973PhDT........17D}.} (Eq.~(42) of Ref.~\cite{1969ApJ...155...89K})
\begin{equation}\label{eq:astrophysical_parameter_Kantowski}
\mathcal{A}\e{K} = \frac{3}{H_0} (1-\bar{\alpha}) \frac{\ev{r\e{S}^2 r\e{c}^{-2}}}{\ev{r\e{S}}},
\end{equation}
which only differs from Eq.~\eqref{eq:astrophysical_parameter} by the powers of $r\e{S}$ in the averages. In a SC model where all the holes are identical, we thus have $\mathcal{A}\e{K}=\mathcal{A}$, but if their masses are distributed according to the same distribution as in Ref.~\cite{Fleury:2014gha}, then $\mathcal{A}\e{K}/\mathcal{A}=1.9$. Although the calculation leading to Eq.~\eqref{eq:astrophysical_parameter_Kantowski} is not fully detailed in Ref.~\cite{1969ApJ...155...89K}, its discrepancy with our result~\eqref{eq:astrophysical_parameter} may be due to different statistical assumptions. In particular, we suspect that Kantowski took into account that bigger SC holes have a larger probability to be encountered by a light beam, whereas we did not---in our approach, holes are randomly placed on the line of sight, irrespective of their sizes. While the former is relevant in an exact SC model, the latter may better correspond to the actual small-scale structure of the Universe.

\subsubsection{A post-Kantowski-Dyer-Roeder approximation}

In \S~\ref{sec:optical_scalars_order1} we have derived the general expression~\eqref{eq:correction_DA_order1} of the correction~$\delta^{(1)}_{D\e{A}}=(\ev{D\e{A}}-D_0)/D_0$ to the mean angular distance with respect to the zero-shear distance~$D_0$---here given by the KDR approximation. With the formula~\eqref{eq:covariance_Weyl_result} for $C_\Weylfoc$ in a SC model, this \emph{post-Kantowski-Dyer-Roeder} (pKDR) term reads
\begin{empheq}[box=\fbox]{equation}\label{eq:linear_correction_DA_z}
\delta_{D\e{A}}^{(1)}
=
-\mathcal{A} \, \Omega\e{m0}
\int_0^z \frac{\dd z_1}{E(z_1)}
\int_0^{z_1} \frac{\dd z_2}{E(z_2)}
\int_0^{z_2} \frac{\dd z_3}{E(z_3)}
\pac{ \frac{\hat{D}_0^2(z_3)}{\hat{D}_0(z_1)\hat{D}_0(z_2)} }^2.
\end{empheq}
with $E(z)\define H(z)/H_0=\sqrt{\Omega\e{m0}(1+z)^3+\Omega_{\Lambda0}}$, and where $\hat{D}_0(z)\define (1+z)D_0(z)$ is sometimes called the corrected luminosity distance, here associated with the KDR distance $D_0$.

\begin{figure}[h!]
\centering
	\begin{subfigure}[b]{0.47\textwidth}
	\centering
	\includegraphics[width=\linewidth]{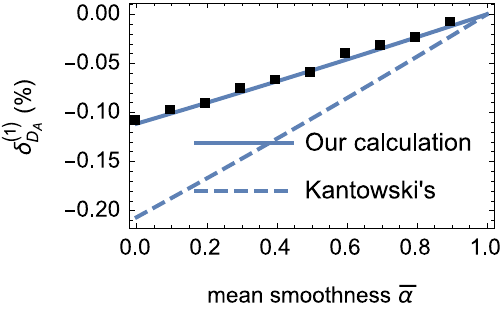}
	\caption{Post-KDR correction $\delta_{D\e{A}}^{(1)}$ as a function of the smoothness parameter~$\bar{\alpha}$, at redshift~$z=1$. Black squares are results from simulations.}
	\label{fig:pKDR_alpha}
	\end{subfigure}
\hfill
	\begin{subfigure}[b]{0.47\textwidth}
	\centering
	\includegraphics[width=\linewidth]{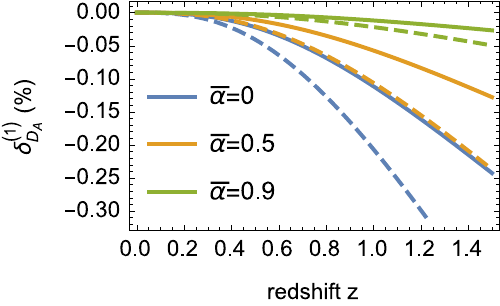}
	\caption{Post-KDR correction $\delta_{D\e{A}}^{(1)}$ as a function of redshift~$z$ for three different smoothness parameters~$\bar{\alpha}=0,0.5,0.9$.}
	\label{fig:pKDR_z}
	\end{subfigure}
\caption{pKDR correction on the angular diameter distance~$\delta_{D\e{A}}^{(1)}\define (\ev{D\e{A}}-D_0)/D_0$, at linear order in Weyl lensing, in SC models made of galaxylike clumps, with $\mathcal{A}_0= 0.5$. Solid lines correspond to our calculations and dashed lines to Kantowski's.}
\label{fig:pKDR}
\end{figure}

Figure~\ref{fig:pKDR} represents $\delta_{D\e{A}}^{(1)}$ as a function of the smoothness parameter~$\bar{\alpha}$ (\ref{fig:pKDR_alpha}) and of the redshift~$z$ (\ref{fig:pKDR_z}), comparing our calculation with the earlier result of Kantowski~\citep{1969ApJ...155...89K}. On Fig.~\ref{fig:pKDR_alpha} are also plotted the results of ray-tracing simulations in SC model, as described in Ref.~\cite{Fleury:2014gha}. Each square represents the average of $(D\e{A}-D_0)/D_0$ over 1000 runs. These numerical results are thus in excellent agreement with the predictions of the stochastic lensing calculations, which proves its efficiency.

The results depicted on Fig.~\ref{fig:pKDR}, namely $\delta^{(1)}_{D\e{A}}\sim 10^{-3}$, confirm that the KDR approximation provides a very good effective description of the angular distance-redshift relation in SC models~\cite{Fleury:2014gha}, at least when galaxy-like clumps are at stake. Nevertheless, since $\delta_{D\e{A}}^{(1)}\propto\mathcal{A}$, this pKDR correction can become very large as the clumps are more compacts; the orders of magnitude given in table~\ref{tab:astro_parameter} suggests that for a SC model made of stars, $\delta_{D\e{A}}^{(1)}\sim 10^7$. This unreasonably large number is a hint that our calculations may break down if too small deflectors are involved. In particular, the infinitesimal light beam approximation---on which both the Jacobi matrix and optical scalar formalisms are based---is not valid for describing the lensing of a star at cosmological distances, which rather requires a microlensing description. See also a discussion by Gunn in Ref.~\cite{1967ApJ...147...61G} on this issue.

\subsubsection{Dispersion of the angular distance}

\begin{figure}[h!]
\centering
	\begin{subfigure}[b]{0.47\textwidth}
	\centering
	\includegraphics[width=\linewidth]{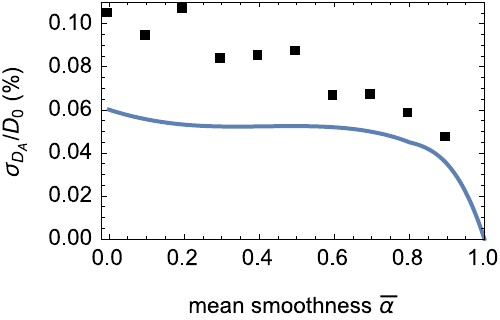}
	\caption{Standard deviation of $D\e{A}$ at redshift $z=1$ as a function of smoothness~$\bar{\alpha}$. Black squares result from ray-tracing simulations and lines from the numerical integration of Eq.~\eqref{eq:varDA_z}.}
	\label{fig:varDA_alpha}
	\end{subfigure}
\hfill
	\begin{subfigure}[b]{0.47\textwidth}
	\centering
	\includegraphics[width=\linewidth]{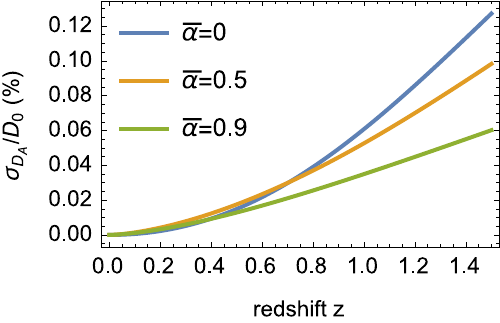}
	\caption{Standard deviation of $D\e{A}$ as a function of redshift~$z$, for three different smoothness parameters, $\bar{\alpha}=0,0.5,0.9$.}
	\label{fig:varDA_z}
	\end{subfigure}
\caption{Standard deviation~$\sigma_{D\e{A}}\define \sqrt{\varDA}$ of the angular distance~$D\e{A}$ in SC models, normalised by the KDR distance $D_0$, as a function of the smoothness parameter~$\bar{\alpha}$ and redshift~$z$.}
\label{fig:variance_DA}
\end{figure}

The general equation governing the variance of the angular distance, $\varDA$, based on second-order calculations in Weyl lensing, has been derived in \S~\ref{sec:variance_angular_distance}. In terms of the redshift, using both Eqs.~\eqref{eq:dx_dv}, \eqref{eq:z_v_FL}, it reads
\begin{empheq}[box=\fbox]{multline}\label{eq:varDA_z}
\Bigg\{
\ddf[3]{}{z} 
+ \pa{\frac{H'}{H} +\frac{6}{1+z}} \ddf[2]{}{z} 
+ \Bigg[ \frac{H''}{H}+\pa{\frac{H'}{H}}^2 
			+ \frac{8}{1+z} \frac{H'}{H}
			+ \frac{6}{(1+z)^2} \\
			- \frac{4\ev{\Ricfoc}}{(1+z)^4 H^2} 
		\Bigg] \ddf{}{z}
+ \frac{2\ev{\Ricfoc}'}{(1+z)^4 H^2} + \frac{ 4C_\Weylfoc - 2C_\Ricfoc }{(1+z)^6 H^3}
\Bigg\} \, \varDA \\
=
\frac{2 D_0^2 C_\Ricfoc}{(1+z)^6 H^3}
+
\frac{6}{(1+z)^6 H^3 D_0^4} \int_0^z \frac{\dd z_1}{(1+z_1)^2 H D_0^2}
\pac{ \int_0^{z_1} \dd z_2 \; \frac{2D_0^4 C_\Weylfoc}{(1+z_2)^2 H}}^2,
\end{empheq}
where a prime denotes here a derivative with respect to $z$. To our knowledge, it is the first time that such a theoretical prediction of the dispersion of the angular distance though a SC model is proposed. This equation is solved numerically, using $\ev{\Ricfoc}=-(3/2)H_0^2 \bar{\alpha}\Omega\e{m0} (1+z)^5$ and the expressions for $C_\Ricfoc$ and $C_\Weylfoc$ derived previously; the output is shown in Fig.~\ref{fig:variance_DA} with, on Fig.~\ref{fig:varDA_alpha}, a comparison with simulated data.

We see that, contrary to its average~$\ev{D\e{A}}$, the standard deviation~$\sigma_{D\e{A}}$ of the angular distance predicted by the stochastic lensing formalism does not fit with the results of ray-tracing simulations. They differ here by a factor $1.7$ for $\bar{\alpha}=0$. We performed a number of consistency checks on both the analytical and numerical sides, and found no errors. It turns out that such a discrepancy between theoretical and numerical results is actually a genuine limitation of our formalism, due to the fact that we modelled Weyl fluctuation by a \emph{Gaussian} noise.

Let us first show that the problem indeed comes from Weyl lensing. Formally, Eq.~\eqref{eq:varDA_z} reads
\begin{equation}
\mathrm{D}_z^3\, \varDA = S_\Ricfoc + S_\Weylfoc,
\end{equation}
where $\mathrm{D}_z$ is a linear differential operator, and $S_\Ricfoc,S_\Weylfoc$ are source functions respectively due to Ricci and Weyl lensing. Contrary to $C_\Ricfoc/C_\Weylfoc$, the ratio $S_\Weylfoc/S_\Ricfoc$ is not necessarily small here; in fact, it is of order unity in the SC model used to generate the results of Fig.~\ref{fig:variance_DA}. A way to tune this ratio---and thus to decide which among Ricci and Weyl fluctuations dominates the dispersion of $D\e{A}$---consists in changing the lower cutoff $b\e{min}=r\e{c}$ of impact parameters in the holes. By virtue of Eq.~\eqref{eq:covariance_Weyl_result}, decreasing $r\e{c}$, i.e. enhancing the compacity of the central clumps, increases $C_\Weylfoc$.

\begin{figure}[t]
\centering
\includegraphics[width=0.47\linewidth]{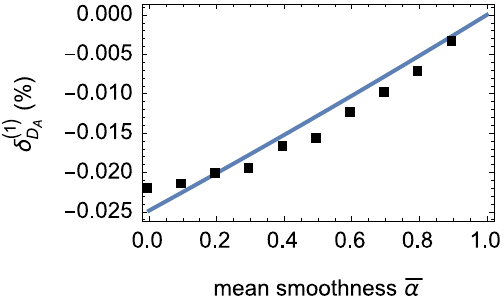}\hfill
\includegraphics[width=0.47\linewidth]{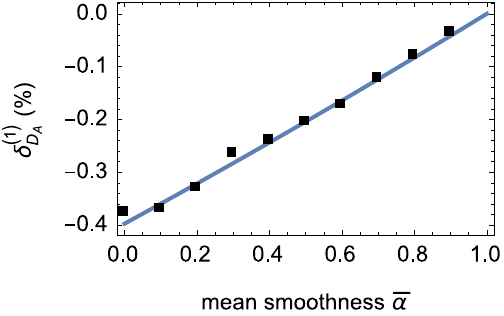}\\[5mm]
\includegraphics[width=0.47\linewidth]{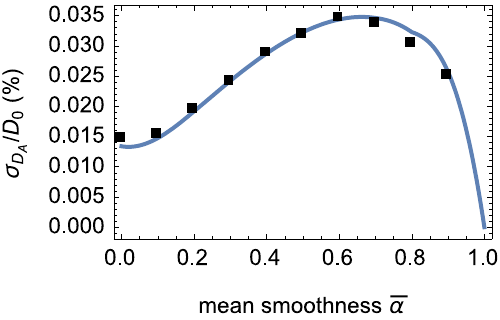}\hfill
\includegraphics[width=0.47\linewidth]{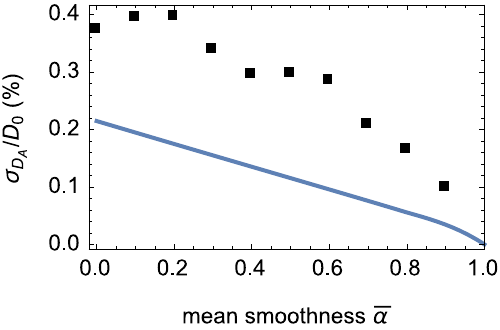}\\[5mm]
\includegraphics[width=0.47\linewidth]{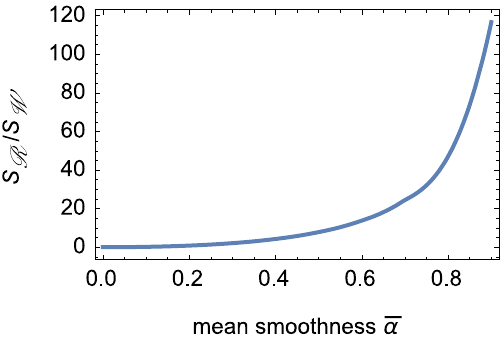}\hfill
\includegraphics[width=0.47\linewidth]{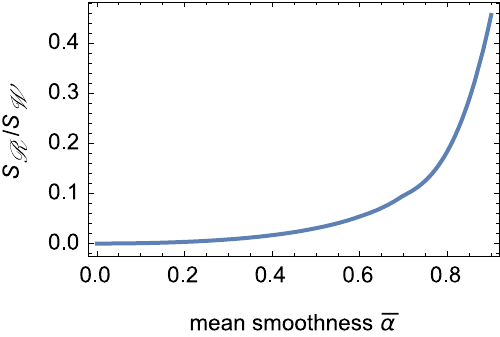}
\caption{pKDR correction to the mean angular distance~$\delta_{D\e{A}}^{(1)}$ (top); normalised standard deviation~$\sigma_{D\e{A}}/D_0$ (middle); and ratio~$S_\Ricfoc/S_\Weylfoc$ between the Ricci and Weyl sources of variance (bottom), as a function of the mean smoothness parameter~$\bar{\alpha}$ at redshift~$z=1$, for SC models with two different clump densities: $\rho\e{c}/8\Leftrightarrow 2 r\e{c}$ (left panel) and $8\rho\e{c}\Leftrightarrow r\e{c}/2$ (right panel), where $\rho\e{c}=3.5\times 10^{-22}\U{kg/m^3}$ is the density used for the previous plots~\ref{fig:pKDR_alpha}, \ref{fig:varDA_alpha}. For a given SC hole, the minimal impact parameter $b\e{min}=r\e{c}$ is thus respectively increased or reduced by a factor $2$ with respect to the previous calculations. As before, squares correspond to the output of ray-tracing simulations, while lines are the predictions of the stochastic lensing formalism.}
\label{fig:reduced_and_enhanced_densities}
\end{figure}

In Fig.~\ref{fig:reduced_and_enhanced_densities}, we compare again the predictions of the stochastic lensing formalism with ray-tracing results, but for two different classes of SC models: with less compact clumps (twice larger for the same mass, left panel); or more compact clumps (twice smaller for the same mass, right panel) than before. We see that the agreement between theory and numerics is now excellent in the first case, where $S_\Ricfoc\gg S_\Weylfoc$, while it is slightly worse than in Fig.~\ref{fig:varDA_alpha} in the second case, where on the contrary $S_\Ricfoc\ll S_\Weylfoc$. The very good agreement regarding $\delta_{D\e{A}}^{(1)}$ in both cases confirms that there are no mistakes in the evaluation of $C_\Weylfoc$

Such results suggest that our modelling of Ricci lensing fluctuations is more accurate than the one of Weyl lensing fluctuations. The weakness does not seem to be related with the $\delta$-correlation hypothesis, because (i) the numerical SC model is constructed so that the properties of two different holes are indeed independent; (ii) the size of the holes is much smaller than the typical evolution scale of $D\e{A}$; and (iii) this hypothesis equally applies to both Ricci and Weyl fluctuations, any deviation from it would therefore be manifest for any value of $S_\Ricfoc/ S_\Weylfoc$, which is not what we observe. 

The Gaussian hypothesis is more questionable. In the standard Langevin description of Brownian motion, the Gaussianity of the random force is justified by the central-limit theorem: during a mesoscopic time interval~$\Delta t$, the Brownian particle is hit by many molecules, and the associated microscopic momentum transfers~$\delta \vect{p}\e{micro}$ \emph{sum} into an effective transfer $\Delta\vect{p}$, whose PDF is therefore well approximated by a Gaussian, whatever the PDF of each $\delta\vect{p}\e{micro}$.

However, while a microscopic Brownian particle undergoes $\sim 10^{20}$ collisions per second, a typical light beam in a SC models encounters only $\sim 10^3$ holes from the source to the observer. The convergence towards central limit must therefore be very efficient for the Gaussian model to be adapted. In the case of Ricci lensing, $\Ricfoc$ simply oscillates between $0$ and $\Ricfoc\e{FL}$; the sum of such a random variable converges quite quickly towards the Gaussian limit, in particular because its support is compact. The case of Weyl lensing is different. One can easily check from the statistical assumptions of \S~\ref{sec:statistical_setup} and the expression~\eqref{eq:effective_Weyl_hole_result} of $\Weylfoc\e{eff}$ that its PDF reads
\begin{equation}
p(|\Weylfoc\e{eff}|) = \frac{2}{3\Weylfoc\e{min}} \pa{ \frac{|\Weylfoc\e{eff}|}{\Weylfoc\e{min}} - \frac{1}{3} }^{-2}
									 [\Weylfoc\e{min}\leq |\Weylfoc\e{eff}|\leq \Weylfoc\e{max}],
\label{eq:PDF_Weyl_True}
\end{equation}
with $\Weylfoc\e{min}=4\pi G\rho_0 (1+z)^5$ and $\Weylfoc\e{max}=\Weylfoc\e{min}[1+2(r\e{h}/r\e{c})^2]/3 \gg\Weylfoc\e{min}$. This PDF thus has a very long algebraic tail, which drastically slows down the convergence towards central limit. This argument is, in our opinion, the most probable explanation of the discrepancy between theory and numerics observed in Fig.~\ref{fig:varDA_alpha}, and of its disappearance in the left panel of Fig.~\ref{fig:reduced_and_enhanced_densities}, where Ricci lensing dominates. This argument shall be reinforced by the results of the next section.

\section{Numerical integration of the Langevin equation}\label{sec:numerics}

The FPK equation, Eq. \eqref{eq:FPK_matrix} for the Jacobi matrix or Eq.~\eqref{eq:FPK_scalar} for the optical scalars, contains all the information necessary to characterise the statistical properties of stochastic lensing, provided this part of lensing can be well approximated by a Gaussian, uncorrelated noise (white noise). However, as mentioned before, it is in general impossible to solve explicitly the FPK equation, and one ought to rely on numerical methods to extract the statistical information available. From the  numerical point of view, solving a partial differential equation is harder than tackling an ordinary differential equation and therefore, it is certainly better to concentrate on the Langevin equation rather than on the FPK equation. In this section, we aim at solving the Langevin equation for the Jacobi matrix, Eq. (\ref{eq:Langevin_matrix}) for a double purpose. First, we wish to show that, in the approximation of a white noise for the Ricci and Weyl lensing, the ray-tracing and analytical results of the previous section are well re-produced by directly solving the Langevin equation. Second, we would like to probe the effect of relaxing the Gaussian approximation: we will show that in the SC model, if the `true' PDF of $\Weylfoc\e{eff}$, Eq.~\eqref{eq:PDF_Weyl_True}, is used, the discrepancy in the fluctuations of $D_{A}$ between the ray-tracing results and the analytical estimates coming from the FPK equation can clearly be attributed to the non-Gaussianity of the noise and the lack of convergence towards the central limit when the number of holes encountered is too small.

\subsection{The stochastic Euler method}

We begin by a short exposition of the numerical discretisation of the general Langevin equation~\eqref{eq:Langevin_general}. For an infinitesimal time step $\dd t$, we can rewrite this equation as
\begin{equation}\label{eq:Langevin_discrete}
\vect{X}(t+\dd t) = \vect{X}(t)+\vect{f}(\vect{X},t) \,\dd t  +   \vect{L}(\vect{X},t)\vect{N}(t) \,\dd t.
\end{equation}
Noting $\dd \vect{B}(t)=\vect{N}(t)\dd t$, this becomes simply
\begin{equation}
\vect{X}(t+\dd t) = \vect{X}(t)+\vect{f}(\vect{X},t)\dd t + \vect{L}(\vect{X},t) \, \dd\vect{B}(t),
\end{equation}
which, after discretisation, gives the Euler approximation to the Langevin equation
\begin{equation}
\label{eq:Euler_Langevin}
\vect{X}(t_{i+1}) = \vect{X}(t_{i})
								+\vect{f}[\vect{X}(t_{i}),t_{i}]\,\Delta t 
								+ \vect{L}[\vect{X}(t_{i}),t_{i}]\,\Delta \vect{B}(t_{i}),
\end{equation}
where we have assumed, for simplicity, a constant time step $\Delta t$. As discussed in Sec.~\ref{sec:from_Langevin_to_FPK}, if the noise~$\vect{N}$ is a white noise, then $\vect{B}$ is a Brownian motion, i.e. its increment~$\Delta\vect{B}$ is a zero-mean Gaussian process with covariance matrix
\begin{equation}
\ev{\Delta\vect{B}(t_{i})\Delta\vect{B}^{\rm T}(t_{j})}
= \vect{\diffusion}(t_{i}) \delta_{ij}\,\Delta t
\qquad \text{(no summation over $i$).} 
\end{equation} 
In practice, simulating one realisation of the process~$\vect{X}(t)$ is thus identical to numerically solving an ordinary differential equation, except that at each time step~$t_i$ the quantity~$\Delta\vect{B}(t_i)$ is randomly picked, according to a Gaussian PDF with variance $\vect{\diffusion}(t_i)\Delta t$, and independently of the other steps. This means that the components of the stochastic term $\Delta\vect{B}$ have fluctuations on the order of $\sqrt{\Delta t}$: the stochastic Euler method only converges as $\sqrt{\Delta t}$, instead of $\Delta t$ for its deterministic counterpart. This limitation will not be a problem in what follows. 

Note also that we can still apply this discretisation if the noise is not Gaussian, but the term $\vect{N}(t)\Delta t$ must then be evaluated from the true PDF of $\vect{N}$. The main caveat, in this case, lies on the fact that the simulation is no longer resolution independent (see \S~\ref{sec:nongaussianity}).

\subsection{Application to the Swiss-cheese model -- Gaussian case}

Let us now turn to our specific Langevin equation for the Jacobi matrix (\ref{eq:FPK_matrix}), in the SC model. Using the relationship between affine parameter and redshift, we can rewrite it as
\begin{equation}\label{eq:Euler_Langevin_SC}
\frac{ \dd\vect{J}}{\dd z} = \frac{1}{H(z)(1+z)^{2}}
												\pac{ \vect{M}(z) \vect{J}(z) + \vect{L}\e{Jac}(\vect{J}) \vect{N}(z) }.
\end{equation}
Discretising this equation with a constant redshift step $\Delta z$ (for simplicity), one gets
\begin{equation}
\label{eq:Euler_Langevin_SC_discretized}
\vect{J}_{k+1} = \vect{J}_{k} + \frac{1}{H(z_{k})(1+z_{k})^{2}} 
													\pac{ \vect{M}(z_{k})\vect{J}_{k} \Delta z 
																+ \vect{L}\e{Jac}(\vect{J}_{k}) \Delta \vect{B}(z_k) },
\end{equation}
with $\vect{J}_{k}\define \vect{J}(z_{k})$, and where the covariance matrix of $\Delta\vect{B}$ involves the diffusion matrix of Eq.~(\ref{eq:Diff_Matrix}) according to
\begin{align}
\ev{ \Delta\vect{B}(z_k) \Delta\vect{B}^{\rm T}(z_k)} &= \vect{\diffusion}(z_k) \Delta z \\
																						&= \mathrm{diag}(C_\Ricfoc,C_\Weylfoc,C_\Weylfoc) \Delta z .
\end{align}
At each time step, the quantity $\transpose{\Delta\vect{B}}=(\Delta B_\Ricfoc, \Delta B_{\Weylfoc_1}, \Delta B_{\Weylfoc_2})$ is obtained by randomly picking $\Delta B_\Ricfoc$ and $\Delta B_{\Weylfoc_A}$ according to a zero-mean Gaussian distribution with variance $C_\Ricfoc\Delta z$ and $C_\Weylfoc\Delta z$, respectively.

\begin{figure}[h]
\centering
\includegraphics[width=0.47\linewidth]{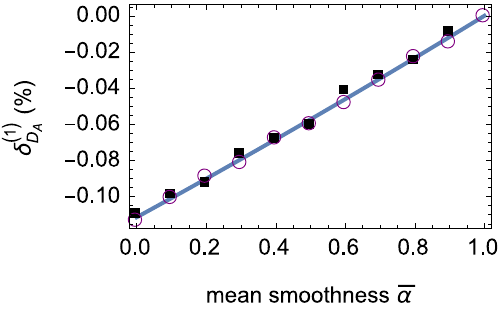}\hfill    
\includegraphics[width=0.47\linewidth]{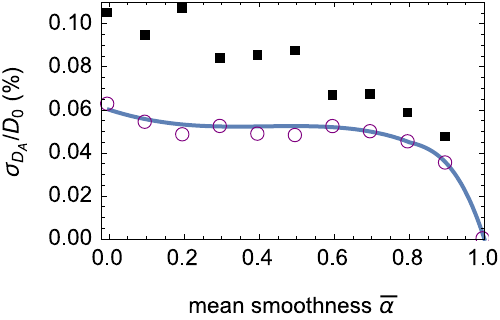}\\    
\caption{Results of the numerical integration of the Langevin equation with a Gaussian noise (empty purple circles), compared with analytical calculations (blue lines) and ray-tracing simulations (black squares) in the SC model.
Left panel (same as Fig.~\ref{fig:pKDR_alpha}): pKDR correction, in percent, to the angular distance at $z=1$, as a function of $\bar{\alpha}$. Right panel (same as Fig.~\ref{fig:varDA_alpha}): fractional dispersion, in percent, of $D\e{A}$ at $z=1$ as a function of $\bar{\alpha}$.}
\label{fig:Langevin_Gauss}
\end{figure}

Using the expressions for $C_{\Ricfoc} (z)$ and $C_{\Weylfoc}(z)$ found in the case of the SC model, Eqs.~(\ref{eq:covariance_Ricci_result}) and (\ref{eq:covariance_Weyl_result}) respectively, we can now integrate numerically the Langevin equation. Results are presented in Fig.~\ref{fig:Langevin_Gauss}, for the same set of parameters as those used in the previous section, i.e. with a standard distribution of galaxy-type holes. Statistical averages are performed over 1000 realisations of $\vect{J}(z)$, for each possible value of $\bar{\alpha}$, each realisation being simulated according to the stochastic Euler method with a redshift step $\Delta z=10^{-4}$. The agreement with the results from the FPK approach is striking, and provides strong support for the analytical expressions found previously. Compared to ray-tracing simulations, the pKDR corrections to $\ev{ D\e{A} }$ are very accurately reproduced, but the dispersion of $D\e{A}$ suffers from the same systematic underestimation as in the previous section. If one could resolve this tension, because the numerical integration of the Langevin equation is much faster than ray-tracing simulations, it would provide an efficient way to estimate statistical quantities.

\subsection{Beyond the Gaussian approximation}\label{sec:nongaussianity}

Now that we have shown that numerically integrating the Langevin equation leads to the same results as the use of the FPK equation in the Gaussian noise limit, we would like to show that the discrepancy between these results and the ray-tracing results stems from non-Gaussianity. Indeed, as discussed in the previous section, the Weyl lensing is very poorly described by a Gaussian noise, since its actual PDF in a SC model presents a long non-Gaussian tail, corresponding to not so rare events during which light rays pass very close to masses and experience significant tidal distortions. In order to probe this effect, in this subsection, we limit ourselves to the case $\bar{\alpha}=0$ in which the Ricci lensing is zero, thus isolating effects due to a pure Weyl lensing. We also use an SC model with one size of holes for simplicity.

We come back to Eq.~(\ref{eq:Euler_Langevin_SC}) but we no longer treat the noise term as the increments~$\Delta\vect{B}$ of a Brownian motion~$\vect{B}$, and replace it by $\Delta\tilde{\vect{B}}$ such that
\begin{align}
\Delta \tilde{B}_{\Weylfoc_1} &= \sqrt{ \frac{3 R\e{h} \Delta z}{(1+z)^4 H(z)}} \,  |\Weylfoc\e{eff}| \cos 2\beta, \\
\Delta \tilde{B}_{\Weylfoc_2} &= -\sqrt{ \frac{3 R\e{h} \Delta z}{(1+z)^4 H(z)}} \,  |\Weylfoc\e{eff}| \sin 2\beta,
\end{align}
where $\beta$ is uniformly distributed within $[0,2\pi]$, while the PDF of $|\Weylfoc\e{eff}|$ is given by Eq.~(\ref{eq:PDF_Weyl_True}). One can check that the above choice ensures that $\ev[2]{ \Delta\tilde{\vect{B}} \Delta\tilde{\vect{B}}^{\rm T}}=\mathrm{diag}(C_\Ricfoc,C_\Weylfoc,C_\Weylfoc) \Delta z$ as before. In other words, the resulting $\Delta\tilde{\vect{B}}$ is a non-Gaussian process whose first two moments match the ones of the Gaussian model. Note that this is somehow artificial, because the noise modelled by $\Delta\tilde{\vect{B}}$ now depends on the resolution~$\Delta z$ used for integrating the Langevin equation. This can be understood as follows. Suppose one solves the Langevin equation with two different resolutions: a low resolution (LR) $\Delta z\e{LR}$, and a high resolution (HR) $\Delta z\e{HR}=\Delta z\e{LR}/n$. During a given interval $[z,z+\Delta z\e{LR}]$, the HR simulation performs $n$ steps, and the effective noise associated with the set of these $n$ steps reads
\begin{equation}
\Delta\tilde{\vect{B}}\e{HR}(z\rightarrow z+\Delta z\e{LR})
=
\sum_{k=0}^{n-1} \Delta\tilde{\vect{B}}\e{HR}(z_k\rightarrow z_{k+1}),
\end{equation}
with $z_k=z+k\Delta z\e{HR}$. Contrary to the Gaussian case, the above sum is \emph{not} equal to $\Delta\tilde{\vect{B}}\e{LR}$, because any random variable does not enjoy the invariance under addition; in particular, for $n\rightarrow\infty$ it becomes Gaussian itself, by virtue of the central limit theorem. 

\begin{figure}[h]
\centering
\includegraphics[width=0.7\linewidth]{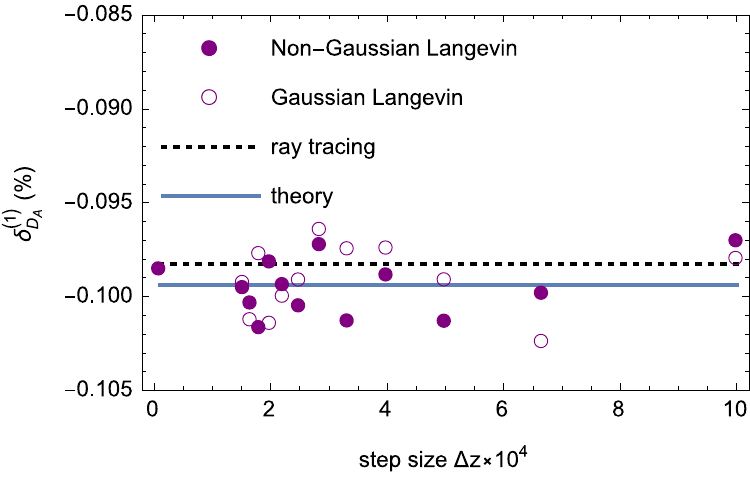}\\    
\includegraphics[width=0.7\linewidth]{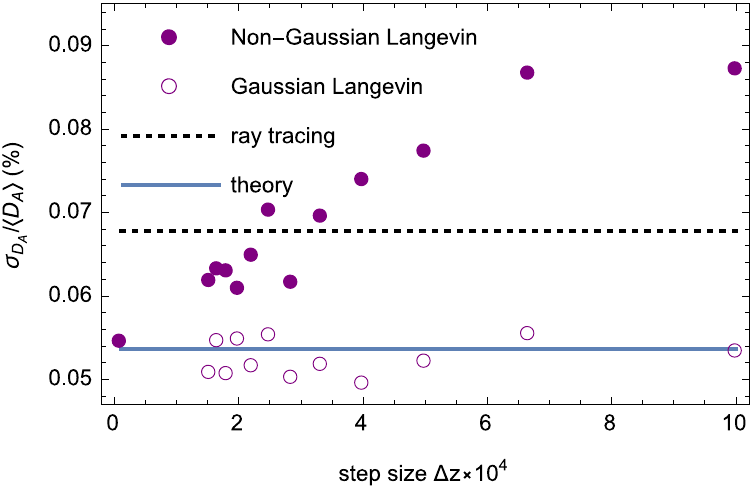}    
\caption{pKDR correction to $\ev{D\e{A}}$ (top panel) and dispersion of $D\e{A}$ (bottom panel) at $z=1$, computed from numerical integration of the Langevin equation~\eqref{eq:Euler_Langevin_SC} with a Gaussian noise~$\Delta\vect{B}$ (empty circles) or a non-Gaussian noise~$\Delta\tilde{\vect{B}}$ (filled circles), as functions of the redshift step~$\Delta z$ used in the Euler method. Each circle is obtained from the statistical properties of a sample of 1000 realisations of $\vect{J}$. For comparison, dotted lines indicate the output of ray-tracing simulations, while the solid lines are the analytical predictions of the stochastic lensing formalism, i.e. given by Eqs.~\eqref{eq:linear_correction_DA_z}, \eqref{eq:varDA_z}. The smoothness parameter of the underlying SC model is $\bar{\alpha}=0$; all holes have the same mass~$M=10^{11}M_\odot$ and density~$\rho\e{c}$.}

\label{fig:NonGauss_Effects}
\end{figure}

We therefore expect the output of numerical integration of the Langevin equation with $\Delta\tilde{\vect{B}}$ (i) to depend on the resolution~$\Delta z$, and (ii) to converge towards the Gaussian case for $\Delta z\rightarrow 0$. This is illustrated in Fig.~\ref{fig:NonGauss_Effects}, where the mean and dispersion of the angular distance at $z=1$, obtained by integrating Langevin equation in the Gaussian and non-Gaussian cases, are plotted as a function of $\Delta z$. We also indicate, for comparison, the analytical and ray-tracing results. A number of comments shall be formulated about those figures. First, all the results on the mean angular distance~$\ev{D\e{A}}$---more precisely, its pKDR correction~$\delta^{(1)}_{D\e{A}}$---are in excellent agreement. It is not the case concerning the dispersion~$\sigma_{D\e{A}}$ of $D\e{A}$. Then, as expected, the Gaussian numerical results coincide with the analytical calculations, as well as the non-Gaussian result for $\Delta z\rightarrow 0$. The latter however depart from the formers as $\Delta z$ increases, and coincides with the ray-tracing results for $\Delta z\approx 2.5\times 10^{-4}$. This particular value can be understood as follows: physically speaking, a non-Gaussian Langevin simulation with redshift step~$\Delta z$ corresponds to a SC model where successive holes are typically separated by $\Delta z$, that is $z/N$ where $z$ is the redshift of the source and $N$ the typical number of holes between the source and the observer. As a matter of fact, with the parameters used for generating Fig.~\ref{fig:NonGauss_Effects}, the average number of holes encountered by a photon is on the order of $3 000$, corresponding to a $\Delta z\sim 3\times 10^{-4}$, which is very close to the value $2.5\times 10^{-4}$ where the ray-tracing and the stochastic non-Gaussian results match.

This confirms our point that, in the SC models investigated here, the typical number of collisions is marginally too small to warrant a treatment of the lensing in terms of a pure white noise, i.e. with a FPK equation. This understanding of the problem provides two ways of escaping from it: (1) dealing with smaller-scale structures; (2) increasing the redshift $z$ of the source. In both situation, the number~$N$ of deflectors, that is the physical resolution of the problem, is increased, which should thus improve the agreement between exact ray-tracing results and the analytical FPK calculations. A quantitative criterion, allowing us to estimate the precision of the FPK approach, remains nevertheless to be determined.

\section{Conclusion}\label{sec:conclusion}

In this article, we proposed a new theoretical framework in which the gravitational lensing caused by the small-scale structure of the Universe is treated as a diffusion process. The Sachs equations governing the propagation of narrow light beams were provided with stochastic components modelled as white noises. We derived the associated Fokker-Planck-Kolmogorov equations for the PDF of the Jacobi matrix and the optical scalars. We used them to deduce (1) the corrections to the mean angular distance due to Weyl lensing, and (2) a differential equation for the dispersion of the angular distance. These results depend on three free functions, namely the mean Ricci lensing~$\ev{\Ricfoc}$ and the covariance amplitudes of Ricci and Weyl lensing~$C_\Ricfoc, C_\Weylfoc$, which need to be specified from a model. As both an illustration and a test of this new formalism, we applied it to Einstein-Straus Swiss-cheese models. The results on $\ev{D\e{A}}$ offer an extension to the Kantowski-Dyer-Roeder approximation, in excellent agreement with numerical simulations. The theoretical predictions for the variance of $D\e{A}$ are however systematically lower than their numerical counterpart. We located the origin of this discrepancy in the actual non-Gaussianity of Weyl lensing, which cannot be captured by the FPK approach. This was confirmed by direct simulations of the Langevin equation with Gaussian and non-Gaussian source terms.

This new approach has the advantage of dealing with small-scale lensing in a mathematically consistent and efficient way, without the need for computationally expensive ray-tracing simulations. It complements the standard description of weak lensing caused by the large-scale structure, allowing for the effect of smaller scales. It is also very flexible, in the sense that it can be applied, in principle, to any model for the distribution of matter on those scales.

The main limitation of our formalism, under its present form, lies in the assumption of Gaussianity. This hypothesis is indeed central in the general derivation of the Fokker-Planck-Kolmogorov equation, on which our main results are based, but it may not hold in the actual Universe, as illustrated on the particular example of Swiss-cheese models. There is, on the mathematics side, active ongoing research on stochastic processes with non-Gaussian noises. Unfortunately no definite standard prescription about how to modify the FPK equation has been established so far, which is the reason why we did not enter into such discussions in the present article. Nevertheless, from a practical point of view, we empirically checked that the Gaussian limit provides good estimations of the lensing quantities, as far as only their mean and variance are concerned: the largest discrepancies are expected to appear for higher-order moments.

In the future, we plan to apply the stochastic lensing framework to more realistic models than the Swiss cheese, in particular for comparing its output with the numerical results of Refs.~\cite{2013PhRvD..88f3004M,2014PhRvD..89b3009Q}. We also intend to include the effect of peculiar velocities, which is not expected to contain major difficulties. On longer terms, we aim at explicitly combining  our formalism with the standard perturbation theory, which would ideally provide a consistent multiscale treatment of cosmological lensing. This will however require to establish a quantitative criterion about the transition scale from one behaviour to the other. Finally, we emphasize the very general character of the approach presented here, which may also be applied to spacetimes with very different symmetries, to treat e.g. the microlensing due to stars in a galaxy, or the effect of a stochastic background of gravitational waves.

\section*{Acknowledgements}

We warmly thank Jorge Kurchan for sharing with us his knowledge and intuition of statistical physics. We also thank George Ellis, Robin Guichardaz, Yannick Mellier, and Cyril Pitrou for discussions. This work made in the ILP LABEX (under reference ANR-10-LABX-63) was supported by French state funds managed by the ANR within the Investissements d'Avenir programme under reference ANR-11-IDEX-0004-02. JL's work is supported by the National Research Foundation (South Africa).

\appendix

\section{Geometric optics in curved spacetime}\label{app:geometric_optics}

This appendix summarises textbook elements about the propagation of light in arbitrary spacetimes, which aims at supplementing the relatively sharp presentation of Sec.~\ref{sec:beams_two_formalisms}. For more general introductions, see Refs.~\cite{Bartelmann:1999yn,1992grle.book.....S,2004LRR.....7....9P,dubook,Sasaki:1993tu}.

\subsection{Description of a light beam}\label{app.A1}

A light beam is a collection of light rays, that is, a bundle of null geodesics $\{v\mapsto x^\mu(v,y^a)\}$ converging at a given event (here taken to be the observation event $O$), where the two coordinates~$(y^a)_{a=1,2}$ label the rays, while $v$ is the affine parameter along them. There is no need for a fourth coordinate because the beam entirely belongs to the lightcone of $O$, which is an isophase hypersurface.

The wave four-vector $k^\mu \define \partial x^\mu/\partial v$ is a null vector field, tangent to the rays $y^a=\mathrm{cst}$. It satisfies the null geodesic equations
\begin{equation}\label{eq.a1}
	k^\mu k_\mu = 0, \qquad \text{and} \qquad k^\nu \nabla_\nu k_\mu = 0.
\end{equation}

Besides, the relative behaviour of two neighbouring geodesics of the bundle, $x^\mu(\cdot,y^a)$ and $x^\mu(\cdot,y^a+\delta y^a)$, is described by their connecting vector $\xi^\mu \define (\partial x^\mu/\partial y^a) \delta y^a$. If the origin $v=0$ of the affine parametrisation of all rays is taken at $O$, then
\begin{equation}\label{eq:orthogonality_separation_wave}
	k^\mu \xi_\mu = 0.
\end{equation}
As soon as the condition~\eqref{eq:orthogonality_separation_wave} is satisfied, the evolution of $\xi^\mu$ along the beam is governed by the geodesic deviation equation
\begin{equation}\label{GDE}
	k^\alpha k^\beta \nabla_\alpha \nabla_\beta \xi^\mu = {R^\mu}_{\nu\alpha\beta} k^\nu k^\alpha\xi^\beta,
\end{equation}
where ${R^\mu}_{\nu\alpha\beta}$ is the Riemann tensor.

\subsection{The Sachs formalism}\label{app.A2}

Consider an observer, with four-velocity $u^\mu$ ($u_\mu u^\mu=-1$), who crosses the light beam. The spatial direction of propagation of the beam, relative to this observer, is defined as the opposite of the direction in which the observer must look to detect a signal. It is spanned by a purely spatial unit vector $d^\mu$,
\begin{equation}\label{eq.a4}
	d^\mu u_\mu = 0, \qquad d^\mu d_\mu = 1,
\end{equation}
such that (remember that we took $k^\mu$ future oriented in this article)
\begin{equation}\label{eq.a5}
	k^\mu =  -\omega (u^\mu + d^\mu),
\end{equation}
where
\begin{equation}\label{eq.a5bis}
	\omega = 2\pi \nu \define u_\mu k^\mu
\end{equation}
 is the cyclic frequency of the light signal in the observer's rest frame. Note that $\dd \ell = \omega \dd v$ is the proper distance (measured by the observer) travelled by light for a change $\dd v$ of the affine parameter. The redshift $z$ is defined as the relative change between the emitted frequency~$\nu_{\rm s}$, in the source's frame, and the observed frequency~$\nu_{\rm o}$, in the observer's frame, that is
\begin{equation}\label{eq.a6}
	1+z \define \frac{\nu_{\rm s}}{\nu_{\rm o}} 
	= \frac{u_{\rm s}^\mu k_\mu(v_{\rm s})}{u_{\rm o}^\mu k_\mu(v_{\rm o})} .
\end{equation}

Suppose that the observer measures the size and the shape of the light beam. For that purpose, he must use (and thus define) a (spatial) screen orthogonal to the line of sight. This screen is spanned by the so-called Sachs basis $(s_A^\mu)_{A\in\{1,2\}}$, defined by
\begin{equation}\label{eq.a7}
	s_{A}^\mu u_{\mu}=s_{A}^\mu d_{\mu}=0,
	\qquad
	g_{\mu\nu }s_{A}^\mu s_{B}^ \nu=\delta_{AB},
\end{equation}
and by the transport property \eqref{eq:transport_Sachs} below. The projections $\xi_A \define s_A^\mu \xi_\mu$ indicate the relative position, on the observer's screen, of the light points corresponding to two neighbouring rays separated by $\xi^\mu$. Thus, it encodes all the information about the size and shape of the beam.

Consider a family of observers $u^\mu(v)$, along the beam, who wants to follow the evolution of the shape of the beam (typically for shear measurements). For that purpose, they must all use the ``same'' Sachs basis, in order to avoid any spurious rotation of the pattern observed on the  screens. This is ensured by imposing that the Sachs basis is a parallel transported as
\begin{equation}\label{eq:transport_Sachs}
	S_{\mu\nu} k^\rho \nabla_\rho s_A^\nu = 0 ,
\end{equation}
where
\begin{equation}\label{eq.a9}
	S^{\mu\nu} = \delta^{AB} s_A^\mu s_B^\nu = g\indices{^\mu^\nu}+u^\mu u^\nu - d^\mu d^\nu
\end{equation}
is the screen projector. The reason why $s_A^\mu$ cannot be completely parallel-transported is that, in general, $u^\mu$ is not\footnote{In fact, it is also possible to choose a family of observers such that the four-velocity field $u^\mu$ is parallel-transported along the beam, without affecting the optical equations~\cite{1992grle.book.....S}. In this case, however, the observers are generally not comoving, and thus have no clear cosmological interpretation.}.

The evolution of $\xi_A$, with light propagation, is determined by projecting the geodesic deviation equation \eqref{GDE} on the Sachs basis. The result is known as the {\em Sachs equation}~\cite{1961RSPSA.264..309S,Seitz:1994xf,1992grle.book.....S},
\begin{equation}\label{eq:Sachs}
	\frac{\dd^2\xi_A}{\dd v^2} = \mathcal{R}_{AB} \, \xi^B,
\end{equation}
where
\begin{equation}\label{eq.a11}
	\mathcal{R}_{AB}={R}_{\mu\nu\alpha\beta}k^\nu k^\alpha s_A^\mu s_B^\beta
\end{equation}
 is the screen-projected Riemann tensor, usually called the {\em optical tidal matrix}. The properties of the Riemann tensor imply that this matrix is symmetric, $\mathcal{R}_{AB}=\mathcal{R}_{BA}$. Note that the position of the screen indices ($A, B, \ldots$) does not matter, since they are raised and lowered by $\delta_{AB}$. In this article, to alleviate the notation, we use bold symbols for quantities with screen indices, and an overdot for derivatives with respect to the affine parameter~$v$. The Sachs equation~\eqref{eq:Sachs} thus becomes
 \begin{equation}\label{eq.a12}
 	\ddot{\vect{\separation}}=\vect{\optical} \vect{\separation}.
\end{equation}

\subsection{Evolution in terms of potentials}

It is interesting to note that the Sachs equation can be reformulated in terms of a potential, as
\begin{equation}
\ddot{\xi}_A = \frac{\partial V}{\partial \dot{\xi}^A},
\end{equation}
with
\begin{equation}
V(\vect{\jacobi}) \define -\frac{1}{2} \xi_A \optical_{AB}\xi_B 
											= -\frac{1}{2} \transpose{\vect{\xi}}\vect{\optical}\vect{\xi}.
\end{equation}
This equally applies to the Jacobi matrix equation $\ddot{\jacobi}_{AB} = -\partial V\e{Jac}/\partial\jacobi_{AB}$, with $V\e{Jac}(\vect{\jacobi}) \define -(1/2) \jacobi_{AB}\optical_{AC}\jacobi_{CB}= -(1/2) \tr\pa{\transpose{\vect{\jacobi}}\vect{\optical}\vect{\jacobi}}$.

Regarding the optical scalars, the Sachs equations can be rewritten as\footnote{As usual, we define the complex derivative as 
\begin{equation}
\frac{\partial{f}}{{\partial\sigma}}=\frac{1}{2}\left(\frac{\partial{f}}{{\partial\sigma_1}}-\i \frac{\partial{f}}{{\partial\sigma_2}} \right),
\end{equation}
for $\sigma=\sigma_1+\i \sigma_2$.}
\begin{equation}
\frac{{\rm d}}{{\rm d}v}\begin{pmatrix}
\nul{\theta}\\ \nul{\sigma} 
\end{pmatrix}
=
-
\begin{pmatrix}
\partial_{\nul{\theta}}\\ \partial_{\nul{\sigma}} 
\end{pmatrix}
V\e{scal}
+
\begin{pmatrix}
\Ricfoc\\ \Weylfoc
\end{pmatrix},
\end{equation}
where
\begin{equation}
V\e{scal}\define \frac{\nul{\theta}^3}{3} + |\nul{\sigma}|^2\nul{\theta},
\end{equation}
and the focusing scalars $\Ricfoc$ and $\Weylfoc$ are here treated as an external force; they could also have been included in the potential according to
\begin{equation}
\tilde{V}\e{scal} \define V\e{scal} - \Ricfoc\nul{\theta} - \Weylfoc\nul{\sigma}.
\end{equation}
Note however that $V\e{scal}$ does not depend explicitly on $v$, while $\tilde{V}\e{scal}$ and $V\e{Jac}$ generally do, because of the presence of $\Ricfoc$ and $\Weylfoc$.

\subsection{Decompositions of the Jacobi matrix}\label{app:Jacobi}

As defined in Sec.~\ref{sec:beams_two_formalisms}, the Jacobi matrix relates the physical separation~$\xi^A(v)$ of two neighbouring rays of a beam at $v$ to their observed separation~$\dot{\xi}^A(0)$, as
\begin{equation}
\xi^A(v) = \jacobi\indices{^A_B}(v) \dot{\xi}^B(0).
\end{equation}
When applied to an observed image, $\vect{\jacobi}(v_\source)$ thus returns the intrinsic physical properties of the source.

\subsubsection{General decomposition}

As any $2\times2$ (nonsymmetric) matrix, the Jacobi matrix has 4 degrees of freedom.  It can be decomposed in a way that highlights the geometrical transformations between the source and the image. First of all, up to frequency factor $\omega_{\obs}$ fixed to $1$ in this article, the determinant of the Jacobi matrix is related to the angular diameter distance as
\begin{equation}\label{eq.defDA}
D\e{A}^2(v)
\define\frac{\text{area of the source at }v}{\text{observed angular size}}
= \det\vect{\jacobi}(v).
\end{equation}
Factorising the determinant, we are left with a $2\times 2$ matrix of determinant $1$, which can be decomposed as the product between a symmetric matrix and the exponential of a symmetric traceless matrix:
\begin{equation}
\vect{\jacobi}
=
D\e{A}
\begin{pmatrix}
\cos\psi & \sin\psi \\
-\sin\psi & \cos\psi
\end{pmatrix}
\exp
\begin{pmatrix}
-\gamma_1 & \gamma_2 \\
\gamma_2 & \gamma_1
\end{pmatrix}.
\end{equation}
The exponential matrix can also be diagonalised by defining $\gamma\geq 0$ and $\thet$ as
\begin{equation}
 (\gamma_1, \gamma_2)=\gamma(\cos2\thet,-\sin2\thet),
\end{equation}
so that
\begin{equation}
\vect{\jacobi}
=
D\e{A}
\begin{pmatrix}
\cos\psi & -\sin\psi \\
\sin\psi & \cos\psi
\end{pmatrix}
\begin{pmatrix}
\cos\vartheta & -\sin\vartheta \\
\sin\vartheta & \cos\vartheta
\end{pmatrix}
\begin{pmatrix}
\ex{-\gamma} & 0 \\
0 & \ex{\gamma}
\end{pmatrix}
\begin{pmatrix}
\cos\vartheta & \sin\vartheta \\
-\sin\vartheta & \cos\vartheta
\end{pmatrix}.\label{eq:decomposition_Jacobi}
\end{equation}
This decomposition shows that, in order to reconstruct the physical properties of a light source from its observed image, one must:
\begin{enumerate}
\item Contract it by a factor $\ex{-\gamma}$ along a direction inclined of $\thet$ with respect to the Sachs basis, and stretch it by a factor $\ex{\gamma}$ along the orthogonal direction. This represent the \emph{net shear}, which preserves the area of the image.
\item Rotate anticlockwise the result by an angle $\psi$.
\item Scale it with $D\e{A}$ to turn angles into lengths.
\end{enumerate}

Note that, by virtue of Etherington's reciprocity relation~\cite{Etherington1933,2004LRR.....7....9P}, which stipulates that the Jacobi matrix obtained by integrating the Sachs equation~\eqref{eq:Sachs} from the observer~$O$ to the source~$S$ or from the source to the observer are opposite and transposed with respect to each other,\footnote{
This can directly be shown from the fact that $\vect{\optical}^{\rm T}=\vect{\optical}$, which implies that, for any two $v_1,v_2$ the function $\vect{C}(v)$ defined by
$\vect{C}(v) \equiv \dot{\vect{\jacobi}}^{\rm T}(v\leftarrow v_1)\vect{\jacobi}(v\leftarrow v_2)
								-\vect{\jacobi}^{\rm T}(v\leftarrow v_1)\dot{\vect{\jacobi}}(v\leftarrow v_2)$
is a constant. Writing $\vect{C}(v_1)=\vect{C}(v_2)$, we conclude that
\begin{equation}
\vect{\jacobi}(v_1\leftarrow v_2) = -\vect{\jacobi}^{\rm T}(v_2\leftarrow v_1) .
\end{equation}
This relation is central to the derivation of the distance duality relation~$D\e{L}=(1+z)^2 D\e{A}$. The latter is however more easily achieved by abandoning the convention~$\omega_\obs=1$. See e.g. Ref.~\cite{thesis_Pierre} for further details.
}
\begin{equation}
\vect{\jacobi}(S\leftarrow O) = -\transpose{\vect{\jacobi}}(O\leftarrow S),
\end{equation}
the net shear $\gamma$ is independent of the sense in which this integration is made.
This is the general nonperturbative generalization of the shear reciprocity relation mentioned in appendix~A of Ref.~\cite{2015arXiv150706590L}.

\subsubsection{Perturbative case}

Usually, when dealing with weak lensing as caused by perturbations with respect to Minkowski or Friedmann-Lema\^{\i}tre spacetimes, one uses that at background level both shear and rotation vanish so that $\bar{\vect{\jacobi}}=\bar{D}\e{A}\,\identity$. The decomposition~(\ref{eq:decomposition_Jacobi}) can then be expanded at first order in $\gamma_1,\gamma_2$ and $\psi$ to get the definition of the {\em amplification matrix} as
\begin{equation}
\vect{\jacobi}=\vect{\amplification}\bar{\vect{\jacobi}} + {\cal O}(2)
\end{equation}
with
\begin{equation}
\vect{\amplification}
=
\begin{pmatrix}
1-\kappa-\gamma_1 & \gamma_2+\psi \\
\gamma_2-\psi& 1-\kappa+\gamma_1
\end{pmatrix},\label{eq:decomposition_amplification}
\end{equation}
where the {\em convergence} is defined as
\begin{equation}
 \kappa\define 1-\frac{1}{2}\tr\vect{\amplification}=\frac{D\e{A}-\bar D\e{A}}{\bar D\e{A}} + \mathcal{O}(2).
\end{equation}
The rotation angle~$\psi$ can be proved to be on the order of $\gamma^2$, and can thus be omitted at linear order, which yields the standard form of the amplification matrix. The decomposition~(\ref{eq:decomposition_Jacobi}) is however much more relevant in nonperturbative cases, or when either the shear or the rotation does not vanish at background level; see e.g. Ref~\cite{Fleury:2014rea}.


\bibliographystyle{JHEP}
\bibliography{bibliography}


\end{document}